\documentclass[12pt,a4paper]{article} 
\pdfoutput=1
\usepackage[utf8x]{inputenc}     
\usepackage{jheppub}
\usepackage{enumerate}
\usepackage{epsfig} 
\usepackage{float} 
\usepackage[caption = false]{subfig}
\usepackage{wasysym} 
\usepackage{mathrsfs} 
\usepackage{amsfonts} 
\usepackage{amsbsy} 
\usepackage{amscd} 
\usepackage{pstricks} 
\usepackage{multirow}
\usepackage{tikz}
\usepackage{color}
\usepackage{slashed}
\usepackage{array}
\usepackage{tablefootnote}
\usepackage{mathrsfs}
\usepackage{nccmath}
\usetikzlibrary{arrows,positioning,shapes.geometric} 
\usepackage{color}
\definecolor{myred}{rgb}{0.6,0,0} 
\definecolor{myblue}{rgb}{0,0.2,0.4}
\definecolor{mygreen}{rgb}{0,0.9,0.1}
\definecolor{hc}{rgb}{.9,0.1,0.7}
\definecolor{hcout}{rgb}{.9,0.7,0.9}
\definecolor{Orange}{rgb}{1.,0.65,0.}

\title{Boosted jet techniques for a supersymmetric scenario with gravitino LSP} 
\author[a, b]{Akanksha Bhardwaj,}  
\author[c]{Juhi Dutta,}
\author[a]{Partha Konar,} 
\author[d]{Biswarup Mukhopadhyaya}
\author[e]{and Santosh Kumar Rai}

\affiliation[a]{Physical Research Laboratory, \\  Ahmedabad - 380009, Gujarat, India}
\affiliation[b]{Indian Institute of Technology, \\ Gandhinagar - 382424, Gujarat, India}
\affiliation[c]{Institut f{\"u}r Theoretische Physik,\\  Universit{\"a}t Hamburg, Luruper Chaussee 149, 22761 Hamburg, Germany}
\affiliation[d]{Department of Physical Sciences, \\
Indian Institute of Science Education and Research Kolkata, \\ 
Mohanpur, 741246, India}
\affiliation[e]{Regional Centre for Accelerator-based Particle Physics, \\
Harish-Chandra Research Institute, HBNI, \\ Chhatnag Road, Jhusi, Prayagraj 211019, India}
\emailAdd{akanksha@prl.res.in}
\emailAdd{juhi.dutta@desy.de}
\emailAdd{konar@prl.res.in}
\emailAdd{biswarup@iiserkol.ac.in}
\emailAdd{skrai@hri.res.in}

\abstract{ Search for compressed supersymmetry at multi-TeV scale, in the presence of a light gravitino dark matter, can get sizable uplift while looking into the associated fat-jets with missing transverse momenta as a signature of the boson produced in the decay process of much heavier next-to-lightest sparticle. We focus on the hadronic decay of the ensuing Higgs and/or $Z$ boson giving rise to at least two fat-jets and $\slashed{E}_T$ in the final state. We perform a detailed background study adopting a multivariate analysis using a boosted decision tree to provide a robust investigation to explore the discovery potential for such signal at 14 TeV LHC considering different benchmark points satisfying all the theoretical and experimental constraints. This channel provides the best discovery prospects with most of the benchmarks discoverable within an integrated luminosity of $\mathcal{L}=200$ fb$^{-1}$. Kinematic observables are investigated in order to distinguish between compressed and uncompressed spectra having similar event yields.
}
\preprint{HRI-RECAPP-2020-004}    


\begin{document}
\maketitle
\section{Introduction}
\label{sec:into}

Improved analysis techniques, especially in the context of the high-luminosity Large Hadron Collider (LHC), are highly desirable in the pursuit of new fundamental physics. Supersymmetry (SUSY) has been one of the front-runner candidates for beyond standard model (BSM) physics for the last few decades, and its search at experiments provides common ground to many non-SUSY searches too. In view of the null results at the Run 1 and Run 2 of LHC, compressed SUSY (cSUSY) \cite{LeCompte:2011cn,LeCompte:2011fh, Martin:2007gf,Martin:2007hn} has gained relevance in its ongoing pursuit, primarily aimed at looking at the elusive scenario of new physics with a significantly degenerate mass spectra. In such scenarios and more specifically in the minimal supersymmetric standard model (MSSM) with the lightest neutralino ($\widetilde{\chi}^0_1$) as the lightest SUSY particle (LSP), the signals are characterized by soft final state objects including low missing transverse momentum ($\slashed{E}_T$) \cite{Martin:2007hn,Martin:2007gf,Martin:2008aw,LeCompte:2011cn,LeCompte:2011fh,Dreiner:2012sh,Bhattacherjee:2012mz,Bhattacherjee:2013wna,Mukhopadhyay:2014dsa,Dutta:2015exw,
Konar:2016ata,Konar:2017oah,Nagata:2017gci}. However, in non-minimal scenarios, the SUSY signals maybe substantially modified in the presence of alternative candidates for LSP and provide valuable probes of detection for the MSSM sector \cite{Dutta:2017jpe,Dutta:2019gox}. In such cases, the SUSY signal is characterised by the presence of hard objects and large $\slashed{E}_T$ in the final state. Typical compressed spectra are not restricted to cSUSY scenarios only and also show up in a variety of other new physics scenarios such as extra-dimensions \cite{Dimopoulos:2014psa,Chakraborty:2017kjq} as well as in extended gauge sectors \cite{Deppisch:2014aga} demanding further phenomenological studies in this context.
   
We focus on compressed SUSY scenarios with a higgsino-like $\widetilde{\chi}^0_1$ (with higgsino fraction $\geq$ 95$\%$) as the next-to-lightest sparticle (NLSP) and a light keV-scale gravitino ($\widetilde{G}$) as the LSP and potential dark matter (DM) candidate. We henceforth refer to this scenario as 'constrained' SUSY ($\mathscr{C}$SUSY). The rest of the spectrum, comprising of the strong and electroweak sparticles, are compressed in mass with respect to the NLSP\footnote{For the rest of the paper, we refer to this compression as the compressed spectra.}. Such a spectrum has previously been studied in the context of MSSM \cite{Dutta:2017jpe} and its extensions \cite{Dutta:2015exw,Dutta:2019gox} at LHC and Tevatron \cite{Matchev:1999ft}. 
In this case, a higgsino-like $\widetilde{\chi}^0_1$ NLSP decays to a Higgs boson or a $Z$ boson along with the $\widetilde{G}$. Therefore the final states arising from the decay of the heavy sparticles lead to 
multifarious diboson ($hh, ZZ, Zh$) signals with large $\slashed{E}_T$.
  
 As the mass scale of new physics extends into the multi-TeV regime new techniques have evolved such as jet substructure techniques \cite{Butterworth:2008iy} have gained importance to study boosted objects. It has been extensively used in various new physics scenarios: vector-like quarks \cite{Sirunyan:2016ipo}, two Higgs doublet models \cite{Bhardwaj:2019mts,Patrick:2016rtw}, little Higgs \cite{Kang:2015nga} and seesaw models 
 \cite{Kang:2015nga,Das:2017gke,Bhardwaj:2018lma}. The di-Higgs channel along with missing transverse energy($\slashed{E}_T$) is explored in reference \cite{Kang:2015nga} using b-tagged jets to reconstruct the Higgs. However, in our current scenario, high $p_T$ $\it{b}$-jets suffer from low reconstruction efficiency.  
 We study the impact of applying boosted techniques to study the prospects of observing $\mathscr{C}$SUSY spectra at the $\sqrt{s}=14 $ TeV run of LHC. We also examine some kinematic observables to distinguish between compressed and uncompressed spectra. 
 The unique points covered in this work are as follows:
 \begin{itemize}
 \item We consider compressed SUSY spectra with a higgsino-like $\widetilde{\chi}^0_1$ NLSP and a light keV gravitino as the LSP and dark matter candidate. The MSSM sector is compressed within $200$ GeV with the NLSP while the NLSP-LSP mass gap is $\mathcal{O}(2 \text{ TeV})$. This ensures the presence of a highly boosted Higgs or $Z$ boson in the final state along with $\slashed{E}_T$.
   
 \item The boosted Higgs or $Z$ boson are studied in the final state containing at least two fat-jets along with $\slashed{E}_T$. A  multivariate analysis is performed using Boosted Decision Tree (BDT) techniques with observables such as N-subjettiness, jet mass and energy correlators used to discriminate between signal and background. The BDT technique shows a clear improvement over conventional cut-based analysis techniques as explicitly demonstrated.

 \item We also discuss possible new signatures complimentary to the hadronic channel. From a preliminary parton level estimate, we observe that such signatures are more likely to be observable at the proposed high energy and high luminosity upgrade of the LHC at $\sqrt{s}=27$ TeV. This provides alternate discovery probes to affirm or exclude the presence of a higgsino-like NLSP.
 \end{itemize}

 The paper is organised as follows: in section \ref{sec:model} we discuss the relevant decays of the higgsino-like $\widetilde{\chi}^0_1$ NLSP. In section \ref{sec:benchmarks}, the current experimental constraints from LHC on the current scenario are discussed and some representative benchmark points satisfying current experimental limits are chosen. The detailed signal and background analysis for the two boosted fat-jets and missing energy is performed and results are presented in section \ref{sec:analysis}. New kinematic observables to distinguish between compressed and uncompressed spectra are discussed in section \ref{sec:distinct}. Section \ref{sec:sumcon} summarises and concludes the work.

\section{Decay properties of a higgsino-like NLSP}
\label{sec:model}  
   
Our focus is on a compressed MSSM sector with the higgsino-like 
$\widetilde{\chi}^0_1$ as the NLSP along with a light $\widetilde{G}$ LSP. For more details we refer the readers to reference \cite{Dutta:2017jpe,Dutta:2019gox}. Here we only 
revisit the relevant decays of the NLSP and the current experimental constraints from 
LHC that dictate our choice of benchmark points.

 The branching ratios of the $\widetilde{\chi}^0_1$ decay are governed by its composition and therefore on the value of the parameters $M_1, M_2, \mu$ and $\tan \beta $ \cite{Martin:1997ns,Meade:2009qv,Covi:2009bk,Matchev:1999ft,Dutta:2019gox}. 
 For a gaugino-like $\widetilde{\chi}^0_1$ NLSP, the obvious decay modes to the $Z \, \widetilde{G}$ and $\gamma \, \widetilde{G}$ are open whereas for the higgsino-like 
 case, its decay to the Higgs mode ($h \, \widetilde{G}$) also opens up.  The relevant partial decay widths of the lightest neutralino in the decoupling limit ($\mu << M_1, M_2$) are  \cite{Martin:1997ns,Meade:2009qv,Covi:2009bk}:
\begin{eqnarray} \nonumber
&&\Gamma(\widetilde{\chi}^{0}_{1} \rightarrow h\widetilde{G}) \propto |N_{14}\cos\beta + N_{13}\sin\beta|^2  \, (M_{Pl}m_{\widetilde{G}})^{-2} \\  \nonumber
&&\Gamma(\widetilde{\chi}^{0}_{1} \rightarrow Z\widetilde{G}) \propto (|N_{11}\sin\theta_W - N_{12}\cos\theta_W|^2 + \frac{1}{2}|N_{14}\cos\beta - N_{13}\sin\beta|^2) \,  (M_{Pl}m_{\widetilde{G}})^{-2}
\end{eqnarray}
where $N_{ij}$ refer to the elements of the neutralino mixing matrix. 
The terms proportional to $N_{14}$ and $N_{13}$ denote the Goldstone couplings to $h/Z$ and $\widetilde{G}$ whereas $\theta_W$ denotes the  Weinberg angle and $\tan \beta =  {v_u}/{v_d}$ is the ratio of the \textit{vev's} $v_u$ and $v_d$ of the two Higgs doublets, $H_u$ and $H_d$, respectively.  Note that 
 for the higgsino-like case there is a huge suppression in branching probability to 
$\gamma \, \widetilde{G}$ mode,
\begin{fleqn}
\begin{eqnarray} \nonumber
 &&\Gamma(\widetilde{\chi}^{0}_{1} \rightarrow \gamma \widetilde{G}) \propto |N_{11}\cos\theta_W + N_{12}\sin\theta_W|^2  \, (M_{Pl}m_{\widetilde{G}})^{-2},   \end{eqnarray}
\end{fleqn}
since the photon mode is governed by the bino and wino components which are suppressed as compared to the higgsino fraction. 

\begin{figure}[!tbh]
\begin{center}
 \includegraphics[scale=0.57]{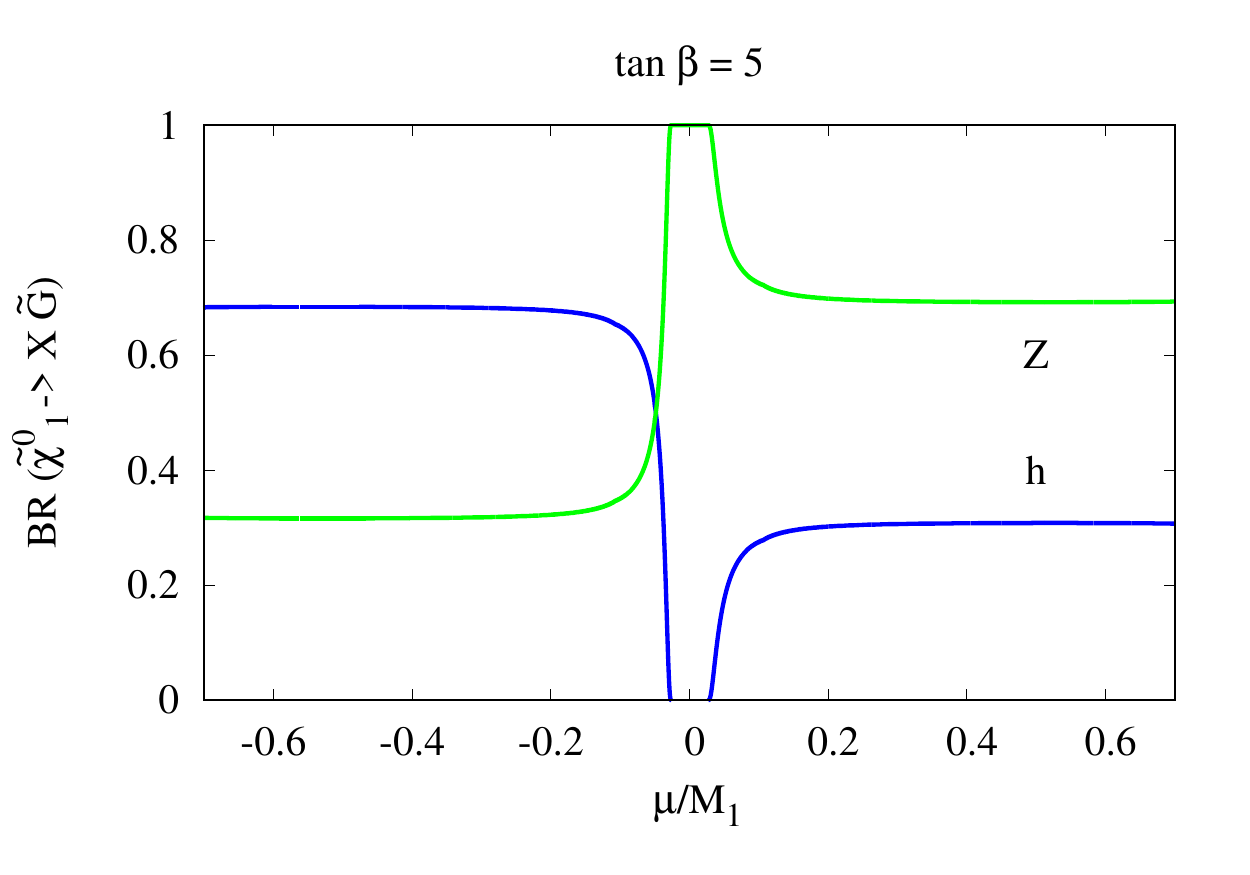}
 \includegraphics[scale=0.57]{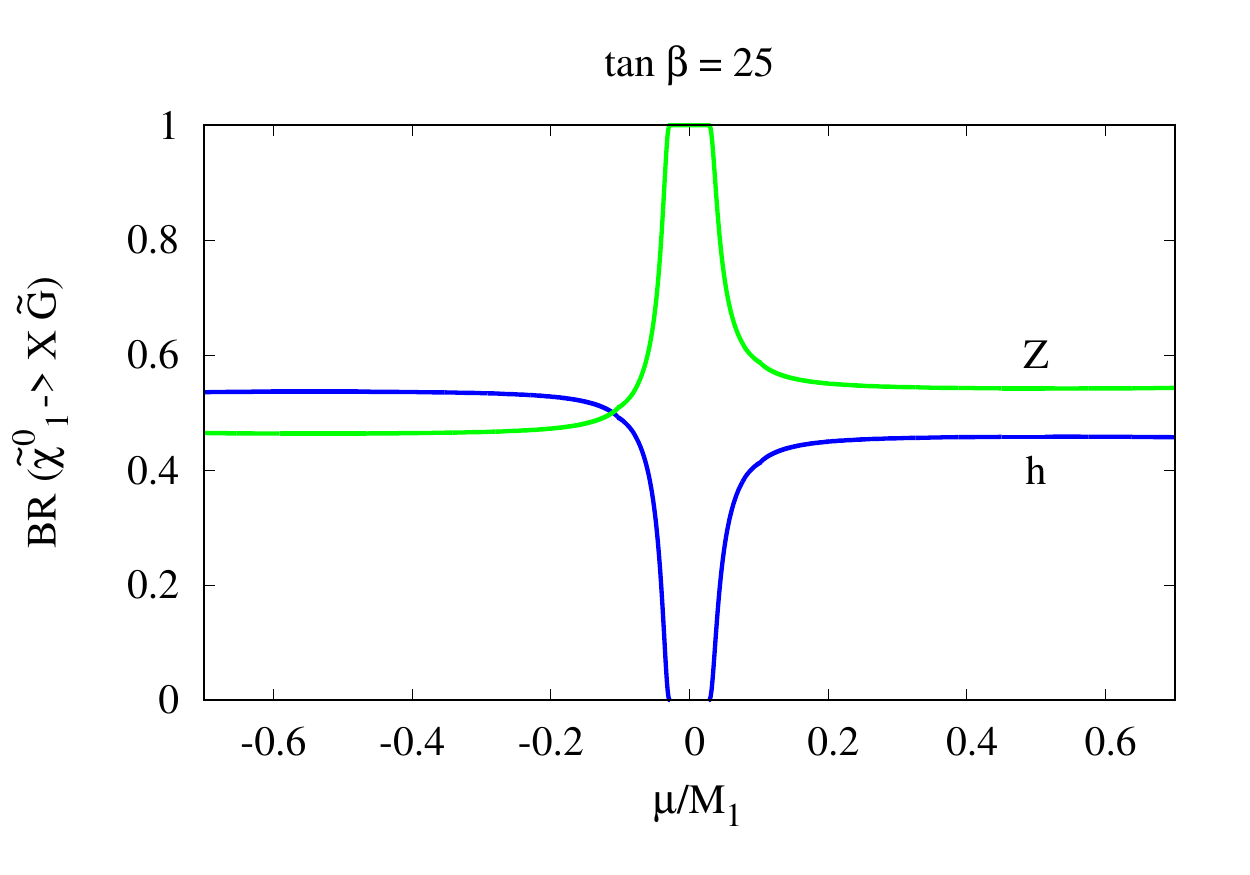}
 \caption{Variation of the branching ratios of the $\widetilde{\chi}^0_1$ NLSP producing the Higgs (blue lines) or $Z$ boson (green lines) as a function of ratio $\frac{\mu}{M_1}$ for fixed values of $M_1, M_2$. All parameters are shown in table \ref{tab:parameters}.  Two plots are for $\tan \beta = 5$ and  $25$ respectively.
 }
\label{fig:par1}
\end{center}
\end{figure}

\begin{table}[tbh]
\begin{center}
 \begin{tabular}{|c|c|c|c|}
\hline
Parameters & $|\mu|$ (TeV)& sign($\mu$)& $\tan\beta$   \\
\hline
Values & 0.2-2.8&$\pm 1$ & 5,25 \\
  \hline
 \end{tabular} 
\caption{Relevant range of the input parameters for the parameter-space scan to study 
the decay probabilities of the lightest neutralino. Other parameters at fixed values which include: $M_1 = 4$ TeV, $M_2 = 4$ TeV, $M_3 =2.9$ TeV, $M_{Q_3} = 2.8$ TeV, $M_{U_3} = 2.8$ TeV, $M_A=3.0$ TeV, $A_t = 3.2$ TeV and $m_{\widetilde{G}}$ = 1 keV. }
\label{tab:parameters}
\end{center}
\end{table}

In figure \ref{fig:par1} we plot the variation of the branching ratios of $\widetilde{\chi}^0_1$ into a Higgs or $Z$ as a function of $({\mu}/{M_1})$. Corresponding fixed values of $M_1, M_2$ and other parameters are listed in table \ref{tab:parameters} where $\mu$ is the higgsino mass parameter, while $M_1$ and $M_2$ are the bino and wino soft mass parameters respectively.  The plots are shown for two values of $\tan \beta = 5, 25$. We have used \texttt{SPheno-v3.3.6} \cite{Porod:2011nf,Porod:2003um} to scan the parameter space.

We observe a gradual increase of the branching into the Higgs with increasing ratio $({|\mu|}/{M_1})$ due to an increase of the higgsino fraction in the NLSP. 
The general features of the plots are summarised below: 
\begin{itemize}
 \item For positive $({\mu}/{M_1})$, the branching ratios to the $Z \, \widetilde{G}$ and 
 $h \, \widetilde{G}$ modes are comparable except in the low $\tan \beta$ regime where 
 the former dominates.
 
 \item For negative $({\mu}/{M_1})$, the $h \, \widetilde{G}$ decay  is greater than 
 $Z \, \widetilde{G}$ decay, primarily in the low $\tan \beta$ regime. 
 
 \end{itemize}
This motivates choice of regions in the parameter space where either decay mode or both 
have branching fractions which are substantial in order to explore the multifarious signal possibilites. Accordingly, we choose the representative benchmarks 
after briefly summarising the relevant experimental constraints in the following section.

\section{Benchmarks}
\label{sec:benchmarks}

\begin{table}[tbh]
	\begin{center}
		\begin{tabular}{|c|c|c|c|}
			\hline
			Final state & Production channels& ATLAS & CMS  \\
			\hline
			$2/3/4 b + \slashed{E}_T$ & $\widetilde{\chi}^{0}_1\widetilde{\chi}^{\pm}_1 ,\widetilde{\chi}^{0}_2\widetilde{\chi}^{\pm}_1,\widetilde{\chi}^{+}_1\widetilde{\chi}^{-}_1, \widetilde{\chi}^{0}_1\widetilde{\chi}^{0}_2$&\cite{Aaboud:2018htj}& \cite{Sirunyan:2018ubx} \\
			$ \ell^+\ell^- + \slashed{E}_T$ & $\widetilde{\chi}^{0}_1\widetilde{\chi}^{\pm}_1 ,\widetilde{\chi}^{0}_2\widetilde{\chi}^{\pm}_1,\widetilde{\chi}^{+}_1\widetilde{\chi}^{-}_1,\widetilde{\chi}^{0}_1\widetilde{\chi}^{0}_2$ & &\cite{Sirunyan:2018ubx}\\
			$\geq 3 \ell +  \slashed{E}_T$ & $\widetilde{\chi}^{0}_1\widetilde{\chi}^{\pm}_1 ,\widetilde{\chi}^{0}_2\widetilde{\chi}^{\pm}_1,\widetilde{\chi}^{+}_1\widetilde{\chi}^{-}_1,\widetilde{\chi}^{0}_1\widetilde{\chi}^{0}_2$ & &\cite{Sirunyan:2018ubx} \\
			$h h + \slashed{E}_T$ & $\widetilde{g}\widetilde{g}$ & & \cite{Sirunyan:2017bsh} \\
			$4 \ell + \slashed{E}_T$& $\widetilde{\chi}^{+}_1\widetilde{\chi}^{-}_1,\widetilde{\chi}^{\pm}_1\widetilde{\chi}^{0}_2$& \cite{Aaboud:2018zeb}& \\
			$\geq2j + \slashed{E}_T$ & $\widetilde{g}\widetilde{g},\widetilde{q}\widetilde{q}$ & \cite{Aaboud:2017vwy} & \cite{CMS-PAS-SUS-19-006}  \\
		        $ b\bar{b} +  \slashed{E}_T$&$\widetilde{\chi}^{0}_2\widetilde{\chi}^{\pm}_1$&\cite{Aaboud:2018ngk} & \\ 
			$ 3 \ell  + \slashed{E}_T$ &$\widetilde{\chi}^{0}_2\widetilde{\chi}^{\pm}_1$ &\cite{Aaboud:2018ngk} & \\ 
			$  \ell^{\pm} \ell^{\pm} + \slashed{E}_T$ &$\widetilde{\chi}^{0}_2\widetilde{\chi}^{\pm}_1$ &\cite{Aaboud:2018ngk} & \\ 
			$2 b + 1 \ell +\slashed{E}_T$ & $\widetilde{\chi}^{\pm}_1 \widetilde{\chi}^0_2$& \cite{ATLAS-CONF-2019-031} & \\
			$\ell^{+}\ell^{-} + \geq 1j  +\slashed{E}_T$ & $\widetilde{\chi}^{\pm}_1 \widetilde{\chi}^0_2$, $\widetilde{l}\widetilde{l}$& \cite{Aad:2019qnd} & \\
			\hline
		\end{tabular}
		\caption{List of the experimental searches from LHC for higgsinos as relevant for our current study with $\widetilde{G}$ LSP.}
		\label{tab:limits}
	\end{center}
\end{table}

Before moving on to choose relevant benchmarks for our current study, we list the currently available constraints from LHC in table \ref{tab:limits}. The current exclusion limits on a light higgsino NLSP and gravitino LSP scenario follow:
 
\begin{itemize} 
 \item  Stringent limits from ATLAS which arise from searches involving multiple $b$-jets along with missing transverse energy ($\slashed{E}_T$)  excluding  $m_{\widetilde{\chi}^0_1} < 380$ GeV for equal branching of the $\widetilde{\chi}^0_1$ into $h \, \widetilde{G}$ and $Z\, \widetilde{G}$ boson. For an increased branching fraction into the Higgs(100$\%$), the mass limits strengthen considerably  excluding  $m_{\widetilde{\chi}^0_1}<890 $ GeV \cite{ATLAShiggsinoMSSM}.
 
\item The CMS Collaboration also sets complementary limits summarized in 
references \cite{CMShiggsinoMSSM, Sirunyan:2017obz, Sirunyan:2018ubx}. Searches involving 
multiple $\it{b}$-jets  and $\slashed{E}_T$ \cite{Sirunyan:2017obz} rule out 
$m_{\widetilde{\chi}^0_1} < 500 $ GeV for 60$\%$ decay of $\widetilde{\chi}^0_1$ into 
$h \, \widetilde{G}$. A combination of searches involving the hadronic search as well 
as multiple leptons and diphotons constrain  $m_{\widetilde{\chi}^0_1}$ up to 700 GeV for equal branching of $\widetilde{\chi}^0_1$ into $h$ and $Z$ along with a $\widetilde{G}$ \cite{Sirunyan:2018ubx}. The exclusion limit improves slightly for the full decay of the $\widetilde{\chi}^0_1$ to the Higgs or $Z$ ($m_{\widetilde{\chi}^0_1} < 750$ GeV).
\end{itemize}
Strongly interacting sparticles are also strongly constrained from LHC searches. A recent study performed using boosted jet techniques in reference \cite{Sirunyan:2017bsh} studies the final state of at least two fat-jets and $\slashed{E}_T$ excluding gluino masses up to 1.8 (2.2) TeV for neutralino LSP mass up to 600 GeV (for $\widetilde{\chi}^0_2$ decaying into Higgs and/or $Z$ boson). This is a relevant constraint for our current work with $\widetilde{G}$ LSP as the same final state was considered thereby imposing strong constraints on the masses of the coloured sparticles.

We choose benchmark points representative of the parameter space allowed by the LHC for a light higgsino-like NLSP scenario with a keV $\widetilde{G}$ LSP.  Our focus is on $\mathscr{C}$SUSY
scenarios as considered in previous studies \cite{Dutta:2015exw,Dutta:2017jpe} with the lightest higgsino-like $\widetilde{\chi}^0_1$ as the NLSP. One also has to accommodate constraints from the observation of a light Higgs in the mass range 122-128 GeV, 
constraints from LEP on the sparticles (primarily  the lightest chargino) as well as 
constraints from flavour physics. The details of such contraints are shown in 
reference \cite{Dutta:2015exw} for the kind of compressed spectra we are interested in. 
The presence of the $\widetilde{G}$ relaxes the dark matter (DM) constraints on the MSSM part of the spectrum with a keV $\widetilde{G}$ DM candidate constituting a warm dark matter candidate \cite{Arvey:2015nra,Covi:2010au,Viel:2005qj,Baltz:2001rq,Boyarsky:2008xj}. We use \texttt{SPheno-v3.3.6}\cite{Porod:2011nf,Porod:2003um} to obtain the 
benchmarks for the current study. We ensure that the benchmarks chosen pass all the relevant experimental searches from Run 1 and Run 2 at the LHC implemented in \texttt{CheckMATE$-$v2.0.26} \cite{Dercks:2016npn}.

Keeping the above constraints in mind, the strongly interacting sector, namely the first and second generation squarks and gluinos, are kept in the mass range 2.4 TeV-3.0 TeV. The third generation squarks are kept heavier than or equal to the first and second generation squarks by choice. In this work we focus on the hadronic signals and choose to keep the electroweak sector heavier than the strong sector. We also focus on a few non-compressed cases to compare the results of our search strategies.
Note that our choice of benchmarks are representative of the parameter space involved. The NLSP decaying to the LSP leads to the presence of either Higgs and/or $Z$ bosons in the final state. Thus the expected final states are $hh+\slashed{E}_T$, $hZ+\slashed{E}_T$ and $ZZ+\slashed{E}_T$, with the light gravitino LSP contributing to the missing transverse momentum ($\slashed{E}_T$). The large NLSP-LSP mass gap ensures that the decay products of the NLSP carry high transverse momentum and hence, a large missing energy in the signal as well. The use of jet substructure techniques will thus be very useful to study the hadronic final state products used to reconstruct the boosted $h/Z$ boson in the final states in order to study the $\mathscr{C}$SUSY spectra. We discuss the analysis techniques and results in section \ref{sec:analysis}.

\begin{table}[tbh]
\scriptsize
	\begin{center}
		\begin{tabular}{|c|c|c|c|c|c|c|c|c|}
			\hline
			Parameters & \textbf{BP1} & \textbf{BP2} & \textbf{BP3} & \textbf{BP4} & \textbf{BP5} & \textbf{BP6} &  \textbf{U1}&\textbf{U2}\\
			\hline 
			$M_1$ &2900& 3000  &3000 &3000 &3500&3500 &2900&2900\\
			$\mu$ &  2340& -2442& 2505&2600& 2812 & 2910  &2390&1000\\
			$\tan \beta$&25 & 25 & 5&25 & 25 & 25 &25 & 25 \\
			$A_t$ & -3200&-3200  & -3300 & -3200&-3200& -3200&-3200& -3200  \\
			$m_A$ &  2500 & 3000&2500&2500 &3000& 3000&3000 & 2500\\
			\hline
			$m_h$ & 124.7& 124.6  &122.1 &124.8 &124.6 & 124.6  &124.7& 124.7 \\
			$m_{\widetilde{g}}$ & 2395.1 & 2494.6  &2609.0 & 2600.9&2999.6 & 2953.3& 3031.7 &3031.7\\
			$m_{\widetilde{q}_L}$ & 2399.1& 2500.9 &2603.5 &2667.7&2983.4 &2961.7 & 2402.1 &2402.2\\
			$m_{\widetilde{q}_R}$ &2398.0 &  2496.7& 2599.3& 2666.4 & 2980.0 &2960.6 &2397.8& 2395.7  \\
			$m_{\widetilde{t}_1}$ &2598.5 & 2612.5&2638.7 & 2612.5& 2893.2 &2929.7 &2606.4& 2587.7  \\
			$m_{\widetilde{t}_2}$ &2787.5 & 2789.8& 2845.9 &2800.2& 3056.0 &3096.5 &2784.7& 2768.2 \\
			$m_{\widetilde{b}_1}$ &2716.1 & 2704.9& 2734.9  & 2726.6&2949.2&2985.6 &2689.2&2690.5 \\
			$m_{\widetilde{b}_2}$ &2781.3&  2790.7 &2789.5 & 2792.3 &3010.1  & 3047.4 & 2784.7& 2722.9\\
			$m_{\widetilde{l}_L}$  &3338.3& 3339.1 &3339.6 & 3339.1&3344.7  &3345.1 & 3338.1&3338.1\\
			$m_{\widetilde{l}_R}$ & 3338.5&3338.8 & 3338.9 &3338.8 &3341.3 &3341.5 &3338.4&3338.5 \\
			$m_{\widetilde{\chi}^{0}_1}$  &2339.5& 2399.9&2498.1 &  2591.0&2809.9 & 2905.1  & 1014.2&2387.3 \\
			$m_{\widetilde{\chi}^{0}_2}$ & -2348.7& -2408.6&-2510.8 &-2603.4&-2817.7 & -2914.0  & -1018.1& -2397.4\\
			$m_{\widetilde{\chi}^{\pm}_1}$ & 2342.7&2402.9&  2502.2 & 2595.1 &2812.7 &2908.2 &1015.9& 2390.8\\
			$m_{\widetilde{\chi}^{\pm}_2}$ &2898.6&  2997.3& 2997.8  &  3004.1 &3485.6& 3486.7  & 2896.2&2897.8\\
			$m_{\widetilde{\chi}^{0}_3}$ &2872.5  &  2972.0& 2971.6&  2974.4& 3463.0 & 3462.0& 2872.5& 2872.6  \\
			$m_{\widetilde{\chi}^{0}_4}$ &2899.0 &  2997.7 & 2998.7&3004.8&3485.9  & 3487.1& 2896.2& 2897.8\\
			\hline
			$\Delta M$ & 59.6 & 101.0 &110.9 &76.7& 189.7 & 56.6 &2017.5& 644.4\\
			$BR(\widetilde{\chi}^{0}_1 \rightarrow Z \widetilde{G})$ & 0.55& 0.55& 0.71&0.56 & 0.55&0.55 & 0.56& 0.55 \\
			$BR(\widetilde{\chi}^{0}_1 \rightarrow h \widetilde{G})$ & 0.45&  0.45 &0.29 &0.44 &0.45 & 0.45&0.44&0.45\\
			\hline
		\end{tabular}
		\caption{List of benchmark points, corresponding parameters and NLSP branching ratios chosen for our study. The mass parameters are in GeV unless specified otherwise. For all benchmarks,  gravitino mass is kept fixed at $m_{\widetilde{G}}=1$ keV.} 
		\label{tab:benchmarks} 
	\end{center}
\end{table}
 
We now discuss the salient features of our benchmark points (BP) as listed in table \ref{tab:benchmarks}. We construct two sets of them as below. While \textbf{BP1}-\textbf{BP6} represent a compressed spectra with narrow mass difference, $\Delta M < 200$ GeV,  \textbf{U1-U2} are for uncompressed spectra having similar yields.
\begin{itemize}

\item \textbf{BP1}-\textbf{BP6}: These represent cSUSY spectra where one has 
comparable branching ratio of the $\widetilde{\chi}^0_1 \rightarrow h \, \widetilde{G}$ and 
$\widetilde{\chi}^0_1 \rightarrow Z \widetilde{G}$ decay modes. The compression parameter ($\Delta M$) which is defined as the difference between the mass of the heaviest colored sparticle (i.e, gluinos or the first and second generation squarks) and the NLSP, varies in the range $\Delta M \simeq 56 -190 $ GeV while $m_{\widetilde{\chi}^0_1} \simeq 2.34-2.91$ TeV. 
	
\item \textbf{U1-U2}: These represent two uncompressed spectra with a lighter NLSP ($m_{\widetilde{\chi}^0_1} \simeq 1.01, 2.39$ TeV) with $\Delta M \simeq 2.02, 0.64$ TeV respectively. 
 \end{itemize}

The different benchmarks involving the compressed spectra vary from one another in the level of mass compression as well as the hierarchical arrangements of the first and second generation squarks and gluinos. For example, \textbf{BP1}--\textbf{BP3}, \textbf{BP5} and \textbf{BP6} have a compressed band involving the strong sector sparticles within 5-10 GeV while \textbf{BP4} accommodates the case where there is a larger mass gap ($ \simeq 67 $ GeV) between the squarks and gluinos. This allows the presence of additional light jets in the latter case as compared to the former ones. 
  
\section{Collider Analysis}
\label{sec:analysis}

\subsection{ Signal topology}
\label{subsec:Sigtop}
In the present study, the lightest neutralino has significant higgsino component which opens up new interesting but challenging channels to study. With the above choice, we can have three interesting final states ($\tilde{\chi_1^0} \tilde{\chi_1^0} \rightarrow hh\tilde{G}\tilde{G}$, 
$ \tilde{\chi_1^0} \tilde{\chi_1^0} \rightarrow ZZ\tilde{G}\tilde{G}$, 
$ \tilde{\chi_1^0} \tilde{\chi_1^0} \rightarrow hZ\tilde{G}\tilde{G}$). It is governed by the benchmarks from table \ref{tab:benchmarks} that the Higgs and the $Z$ boson will be highly boosted and the total hadronic activity of the decay of $h/Z$ can be captured in a large radius jet (fat-jet of radius $R$), which will be directed by the relation \cite{Shelton:2013an}
\begin{equation}
\label{eq:fatR}
R \sim \frac{2 M^{h/Z}}{P_T^{h/Z}}.
\end{equation} 
As shown in table \ref{tab:benchmarks} the mass of neutralino ($\widetilde{\chi}^0_1$) lies in the range of 2-3 TeV. In this case, a Higgs tagger based on b-tagging techniques deteriorates its efficiency \cite{CMS-PAS-BTV-15-002}. In this process, we also lose a sufficient number of events when ($\widetilde{\chi}^0_1$) is decaying to $Z$ boson. To overcome this issue we propose to capture the Higgs and $Z$ candidate using 2-prong finder tagger which is based on the radiation pattern inside the fat-jet. We utilize the jet substructure techniques to identify $h/Z$ candidate by looking for the following signal topology
\begin{center}
$pp \rightarrow $   2 \texttt{CA8} Fat-jets ($J$) +  large $\slashed{E}_T$,
\end{center}
where \texttt{CA8} represents the jets clustered with Cambridge-Aachen algorithm with R = 0.8. The choice of R is decided by the relation given in \ref{eq:fatR} such that both the Higgs and $Z$ boson decay products can be captured with the given cut on the fat-jet momentum ($P_T > 300$ GeV).  However, for similar transverse momentum the fat-jets originating from the Higgs boson have a larger radius then the fat-jets originating from the $Z$ boson. Later we utilize 2-prong finder algorithms like N-subjettiness and energy correlation function (ECF) to tag the Higgs or $Z$ like fat-jets.

\subsection{ Backgrounds}
\label{subsec:bg}
The  major contribution to the background comes from the following Standard Model processes. Corresponding cross sections as used in present analysis are listed in table~\ref{tab:Backgrounds} with  the  order  of  QCD  corrections.

\begin{table}[tbh]
    \centering
    \renewcommand{\arraystretch}{1.5}
    \begin{tabular}{|c|c|}
        \hline
        Background process & cross section (pb) \\
        \hline
        $Z$ + jets [N$^2$LO]& $6.33 \times 10^4$  \cite{Catani:2009sm,Balossini:2009sa} \\ 
        \hline
        $W$ + jets  [NLO]  & $1.95 \times 10^5$    \cite{Alwall:2014hca}\\ 
        \hline
        $Single-top$ $ (tW, tj$ and $tb)$ [N$^2$LO]&  83.1 , 12.35,   248.0  \cite{Kidonakis:2015nna}  \\
        \hline
        $Diboson (ZZ, WW ,ZW)$ + jets [NLO] &  17.72, 124.31,  51.82 \cite{Campbell:2011bn} \\
        \hline
        $t\bar{t}$ + jets [N$^3$LO] &  988.57  \cite{Muselli:2015kba}   \\
        \hline
    \end{tabular}
    \caption{The cross sections for the background processes used in this  analysis are shown with  the  order  of  QCD  corrections provided in brackets.}
    \label{tab:Backgrounds}
\end{table}
\begin{itemize}

\item $Z \rightarrow \nu\bar{\nu} $ +jets turns out to be the most dominating background due to large missing transverse momentum and high fake rate of QCD fat-jets as h/$Z$ jets. 

\item $W \rightarrow l\nu $ +jets contributes to the SM background processes when the lepton is misidentified.  Then the dynamics are the same as $Z$+jets. Due to the large cross-section, these processes contribute significantly.  

\item $VV+jets$: Diboson production in three different channels, such as,  $W_h W_l, W_h Z_{\nu\bar{\nu}}$, and $Z_h Z_{\nu\bar{\nu}}$. Here the $V_h$, $V_l$ and $V_{\nu\bar{\nu}}$ denotes the hadronic, leptonic and invisible decay modes respectively of $W/Z$ bosons. The diboson process has almost similar signal topology but contributes as a subdominant background due to its low cross-section.

\item Single-top production: Among the three different productions of the single top $(tW, tj$ and $tb)$ the main contribution comes from single top associated with W. 

\item $t\bar{t}$ decaying semi-leptonically has the missing transverse momenta from one of the $W$ decaying leptonically and the possible source for fat-jets is either one of the W decaying hadronically or mistagged b-jets.
\end{itemize}

We additionally compute the contributions from the triboson and QCD multijet background which is rendered negligible because of high $\slashed{E}_T$ and two hard fat-jet criteria.

\subsection{  Simulated events and Data sample}
We have generated the $\mathscr{C}$SUSY mass spectrum using {\tt SPheno-v3.3.6}. All the events are generated using {\tt Madgraph5 (v2.6.5)} \cite{Alwall:2014hca} at leading order (LO) followed by {\tt Pythia} (v8) \cite{Sjostrand:2006za} for showering and hadronization. To incorporate detector effects events are passed through {\tt Delphes-v3.4.1} \cite{deFavereau:2013fsa} using the default CMS card. Delphes tower are used as an input for fat-jet clustering. Fat-jet are reconstructed  using the Cambridge-Aachen algorithm \cite{Dokshitzer:1997in} with radius parameter $R = 0.8$, as implemented in the {\tt Fastjet}-v3.3.2 \cite{Cacciari:2011ma}. The minimum $p_T$ for fat-jet is required to be 300 GeV. We use {\tt ROOT 5} \cite{Brun:1997pa} for the baseline event selection. The final multivariate analysis (MVA) is performed using Boosted Decision Tree (BDT), as implemented in toolkit for Multivariate Analysis {\tt TMVA}\cite{Hocker:2007ht}.
The events used in the multivariate analysis are selected after the following baseline cuts which are designed for the signal topology discussed in section \ref{subsec:Sigtop}.\\

\noindent
{\bf Baseline selection criteria }
 \begin{itemize}
   \item  We veto the events if any lepton with $p_T > $ 10 GeV lies in the central psuedorapidity range $|\eta| < $  2.4.
 
 	\item  We select the events with at least two Cambridge-Aachen fat-jets of radius parameter = 0.8 and  with minimum transverse momentum $p_T$ = 300 GeV.
 
  \item To overcome the effect of jet mismeasurement contributing to missing transverse momenta both the fat-jet should satisfy the criteria of $|\Delta \phi(J , \slashed{E}_T)| > 0.2 $.
 	 	
 \item  The signal has large missing energy hence we select the events with $\slashed{E_T}$ greater than 100 GeV.

\end{itemize}

\subsection{Multivariate analysis}
\label{sec:mva}
We perform the collider study using a multivariate analysis (MVA) employing the Boosted Decision Tree (BDT) algorithm. The multivariate analysis outperforms the cut-based analysis since a cut-based analysis can select only one hypercube as the signal region of phase space, whereas, the decision tree can split the phase space into a large number of hypercubes.  Each of these hypercubes is then identified as either a `signal-like' or a  `background-like' tree. Then a non-linear boundary is created in hyperspaces to segregate the signal and background.

\begin{figure}[!t]
\centering
  \subfloat[]{\label{fig:PT_J0}\includegraphics[scale=0.25]{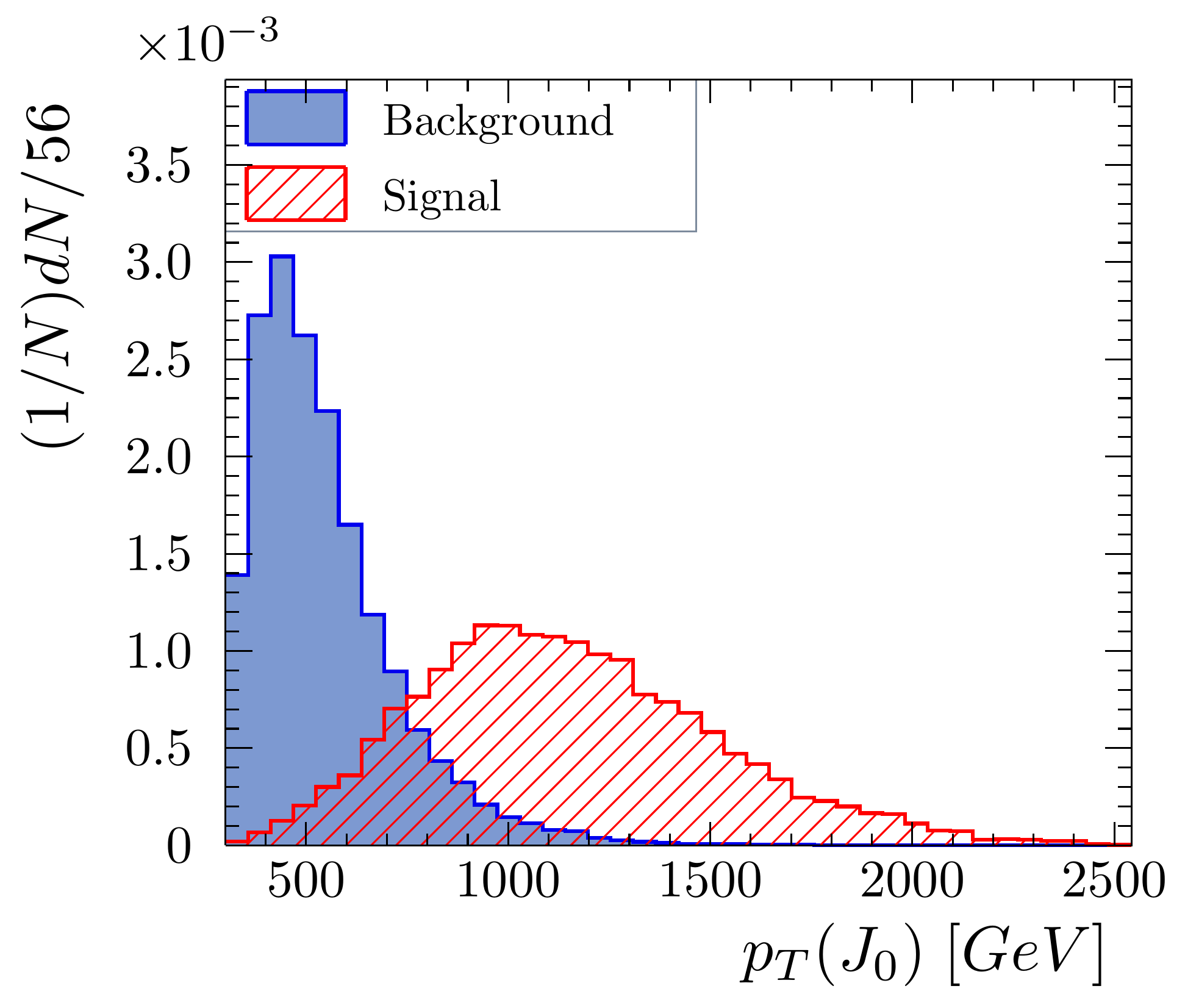}}~
   \subfloat[]{\label{fig:Delta_R_J0_J1}\includegraphics[scale=0.25]{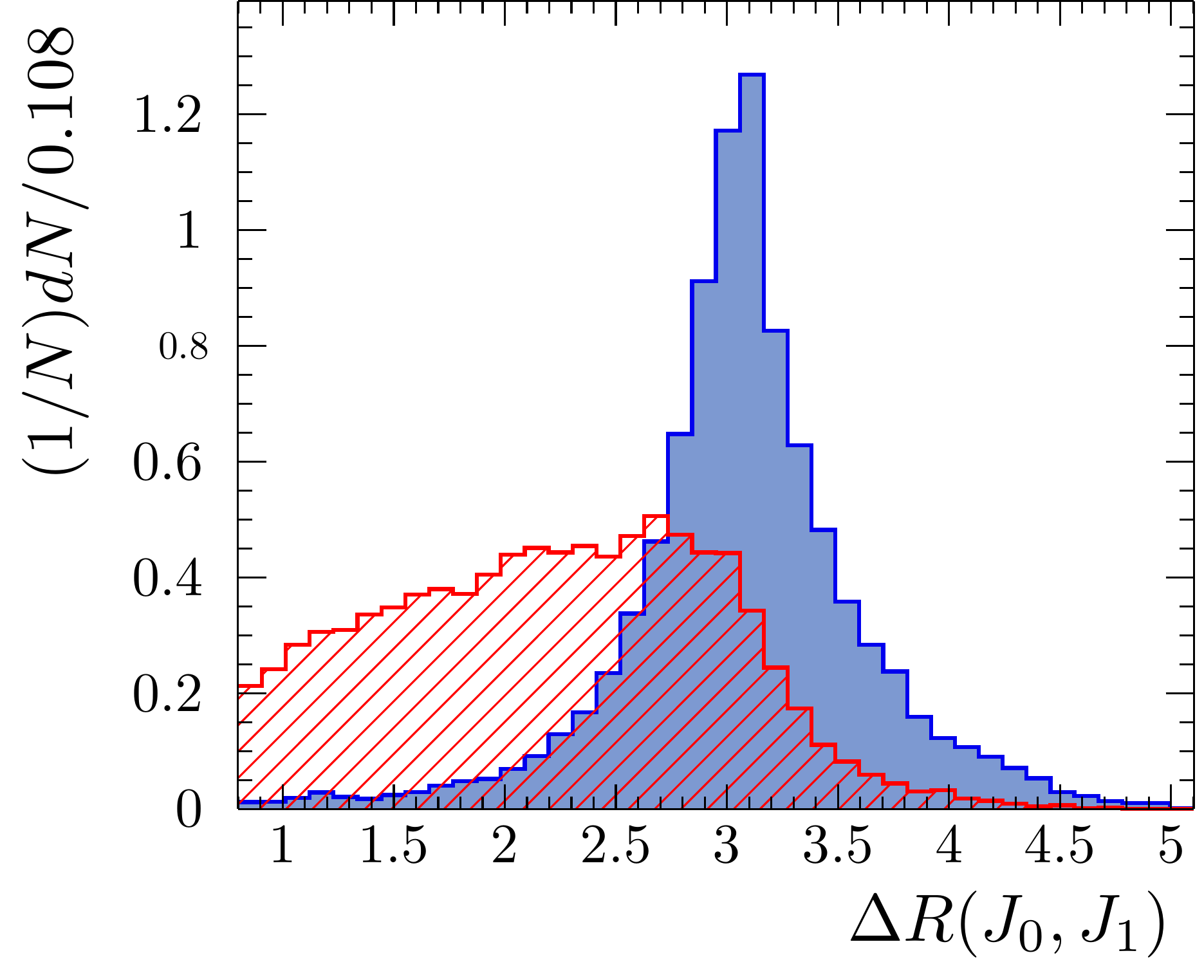}}~
   \subfloat[]{\label{fig:MET}\includegraphics[scale=0.25]{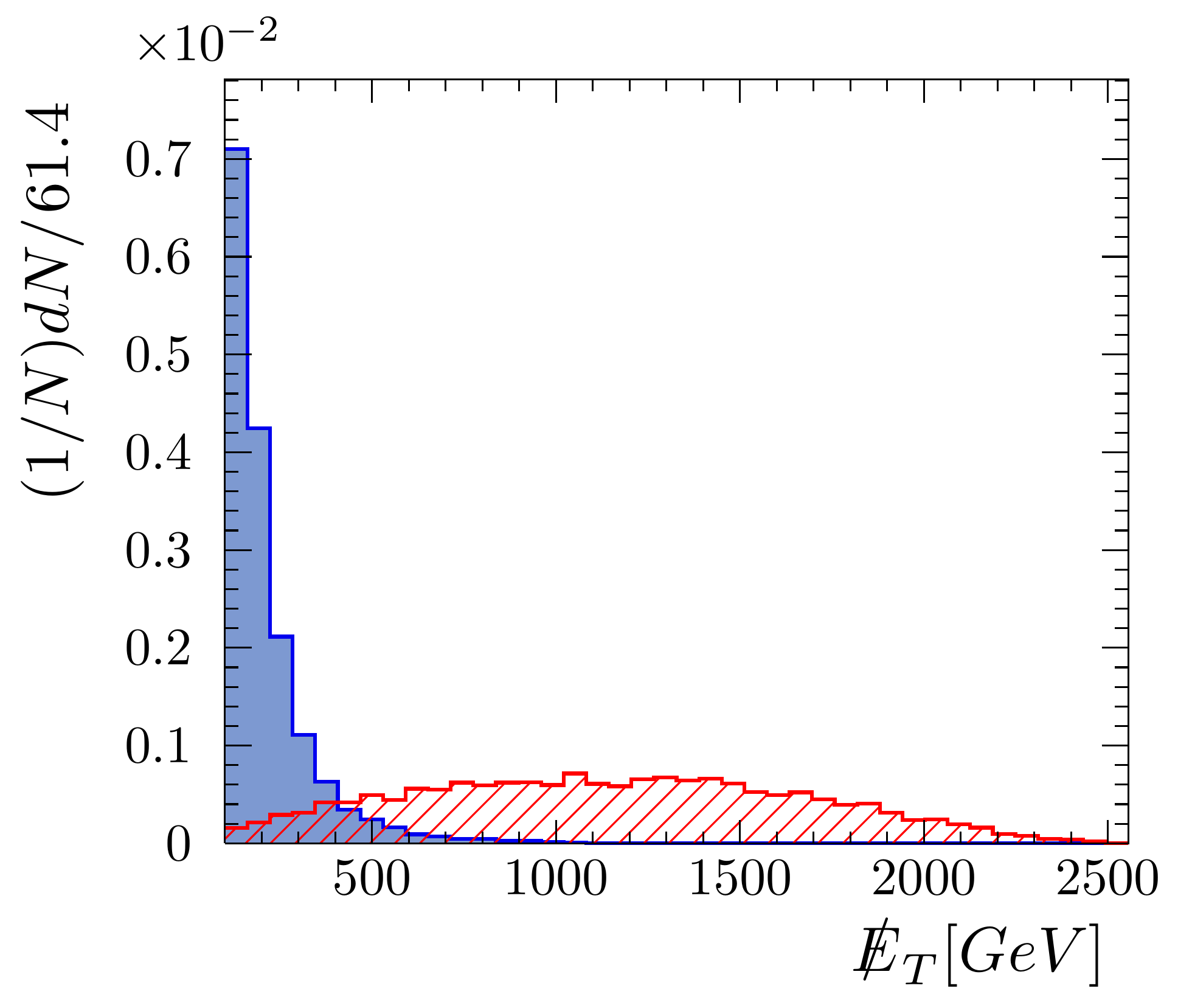}}\\
     \subfloat[]{\label{fig:Delta_Phi_J0_MET}\includegraphics[scale=0.25]{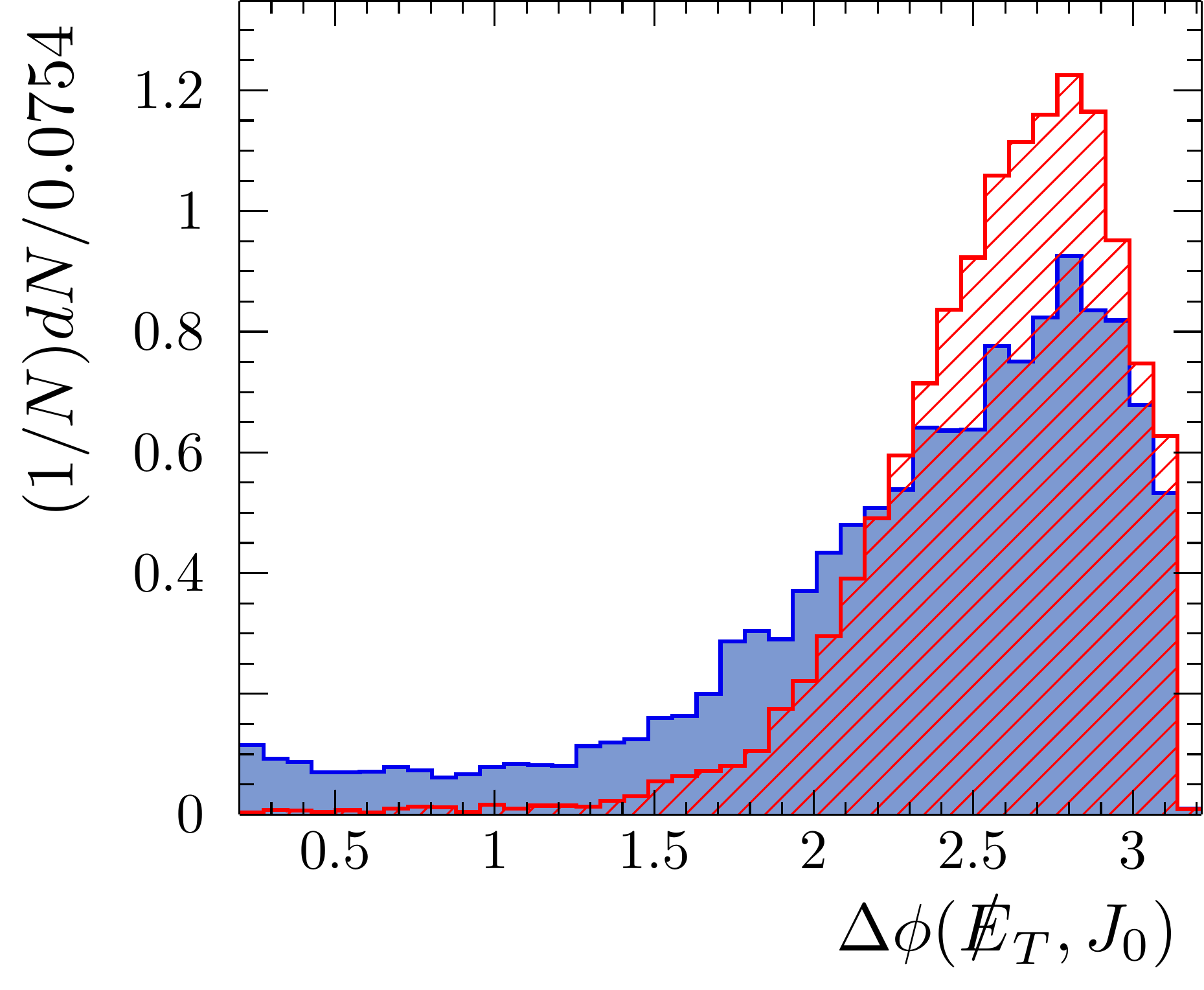}}~
   \subfloat[]{\label{fig:Delta_Phi_J1_MET}\includegraphics[scale=0.25]{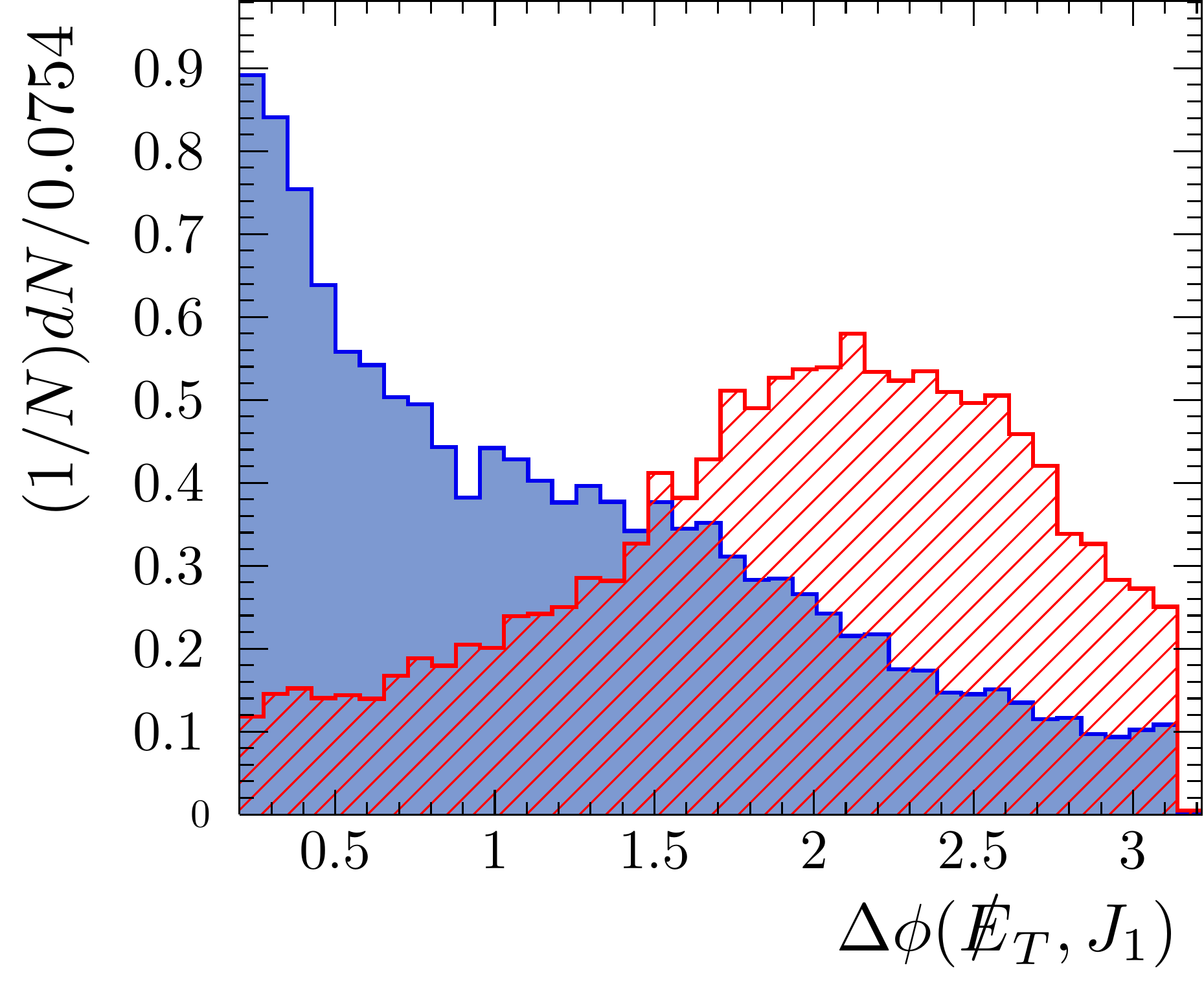}}~
    \subfloat[]{\label{fig:Meff}\includegraphics[scale=0.25]{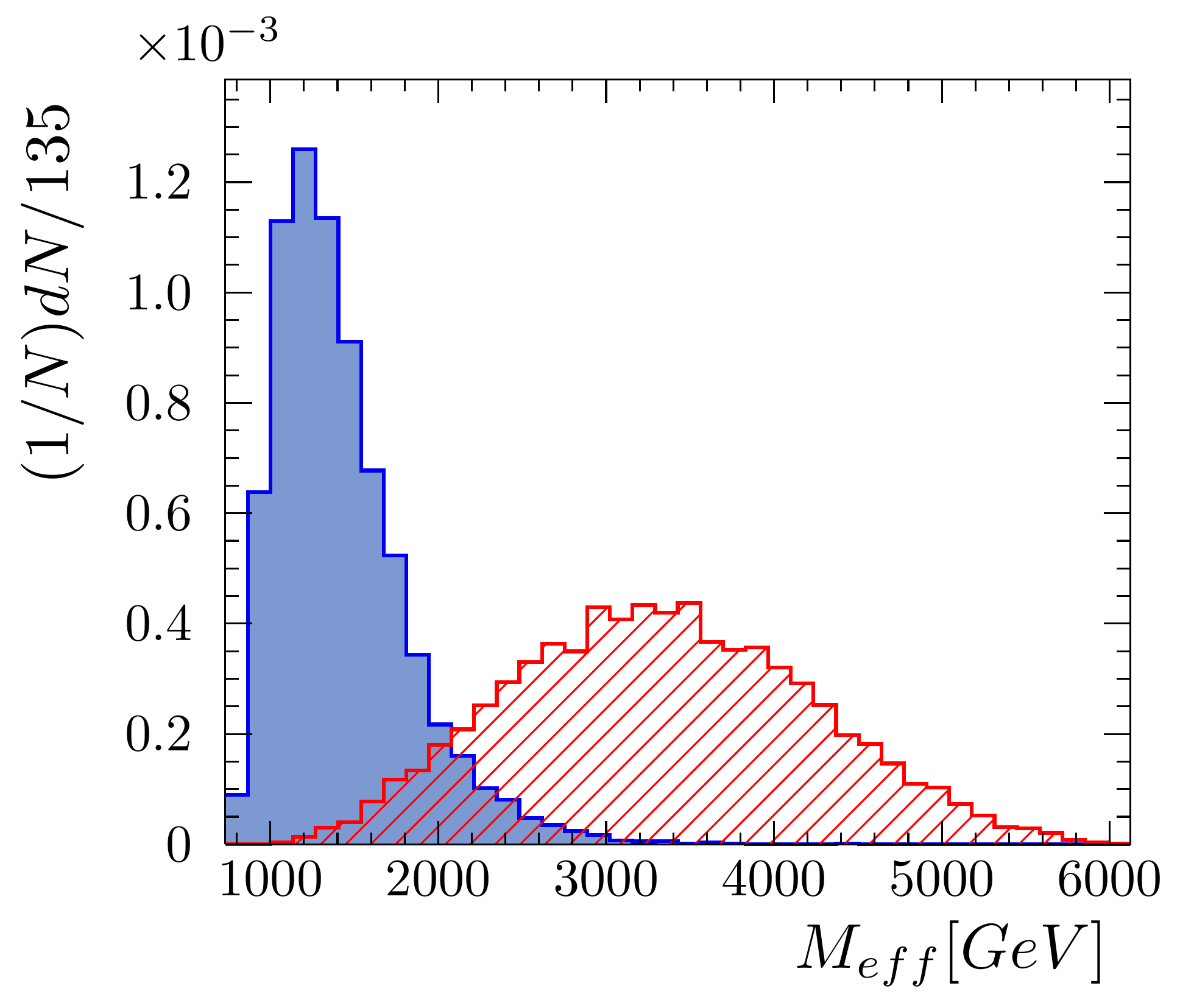}}\\
  \caption{Normalized distributions of the  basic input variables related to two reconstructed fat-jets $J_i$ and missing transverse energy $\slashed{E_T}$  at the LHC ($\sqrt{s}=14$ TeV) used in the MVA for the signal (red) and the background (blue). Signal distributions are obtained for benchmark point {\bf BP1} and the background includes all the dominant backgrounds.}
 \label{fig:4_1}
\end{figure}

\begin{figure}[tbh]
\centering
   \subfloat[]{\label{fig:MJ0}\includegraphics[scale=0.25]{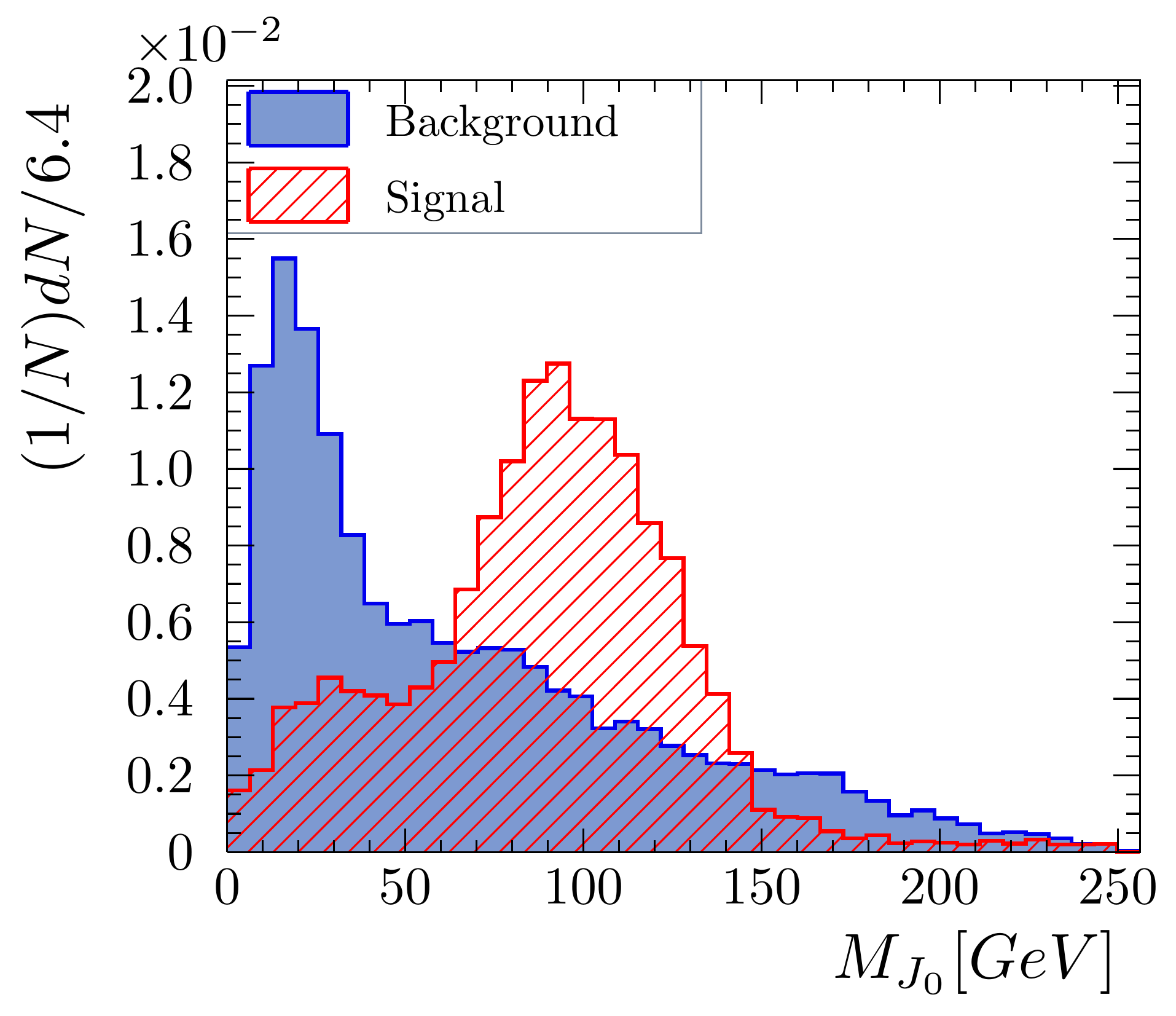}}~
  \subfloat[]{\label{fig:MJ1}\includegraphics[scale=0.25]{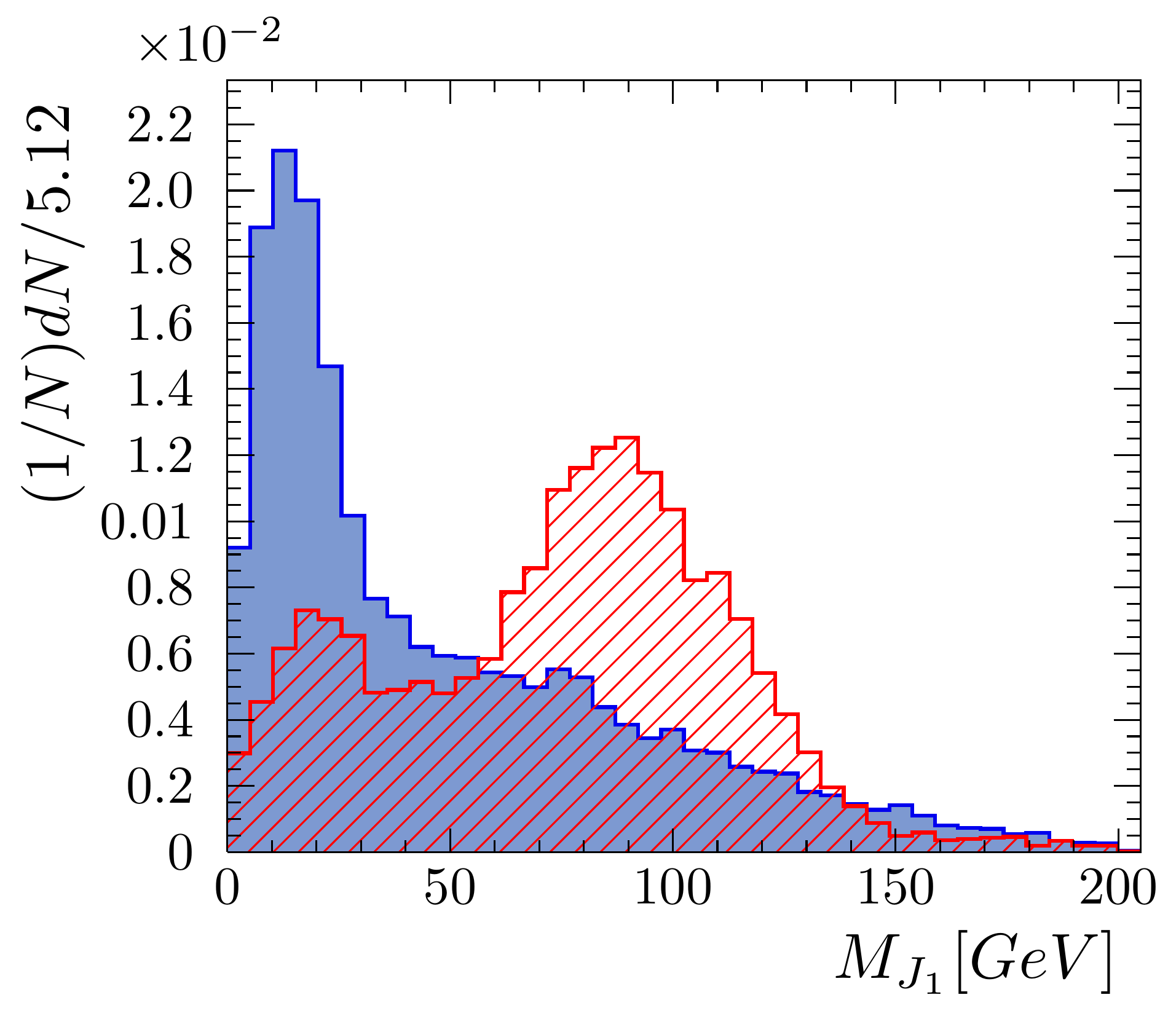}}~
  \subfloat[]{\label{fig:C22_J0}\includegraphics[scale=0.25]{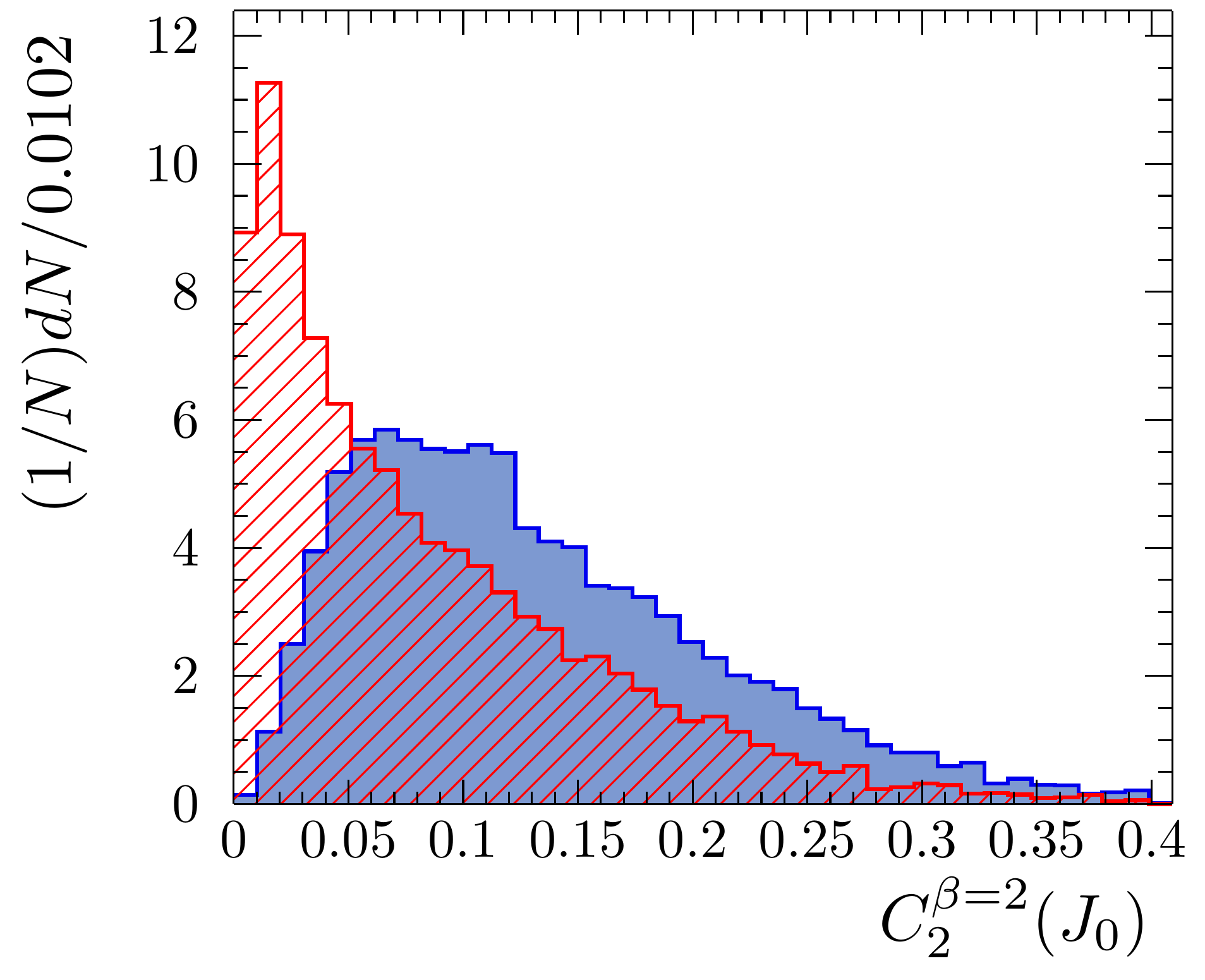}}\\
   \subfloat[]{\label{fig:C22_J1}\includegraphics[scale=0.25]{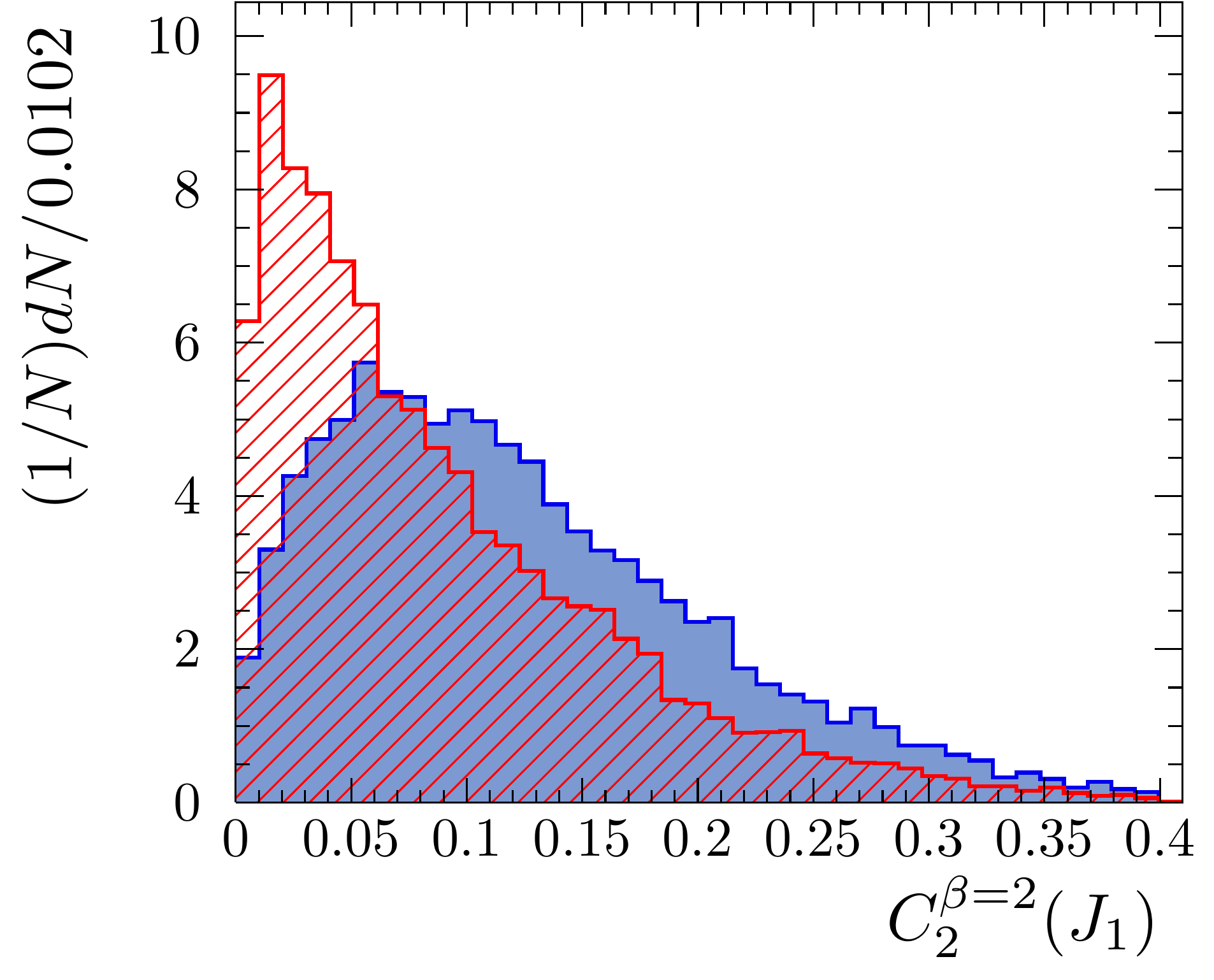}}~
  \subfloat[]{\label{fig:Tau21_J0}\includegraphics[scale=0.25]{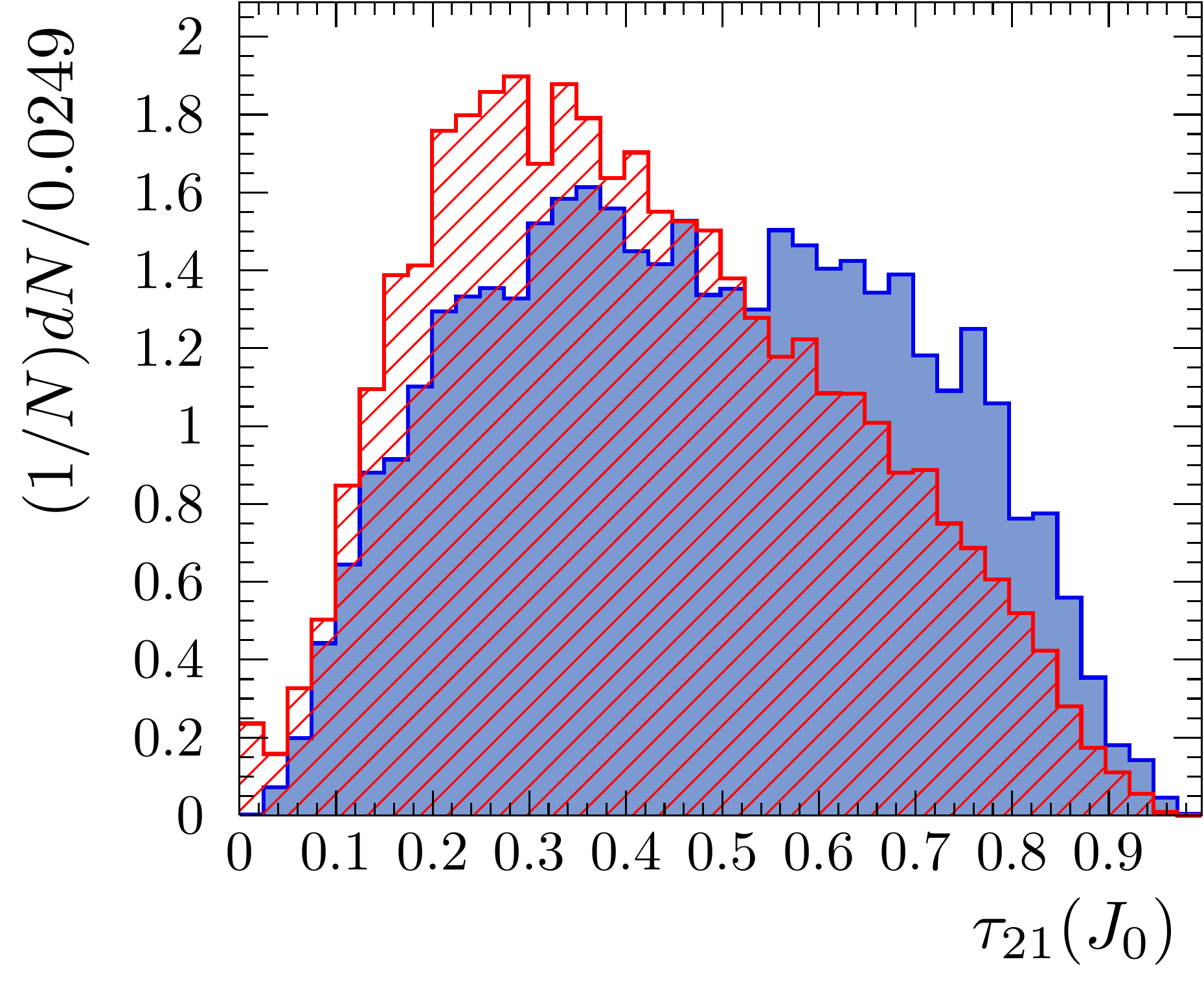}}~
  \subfloat[]{\label{fig:Tau21_J1}\includegraphics[scale=0.25]{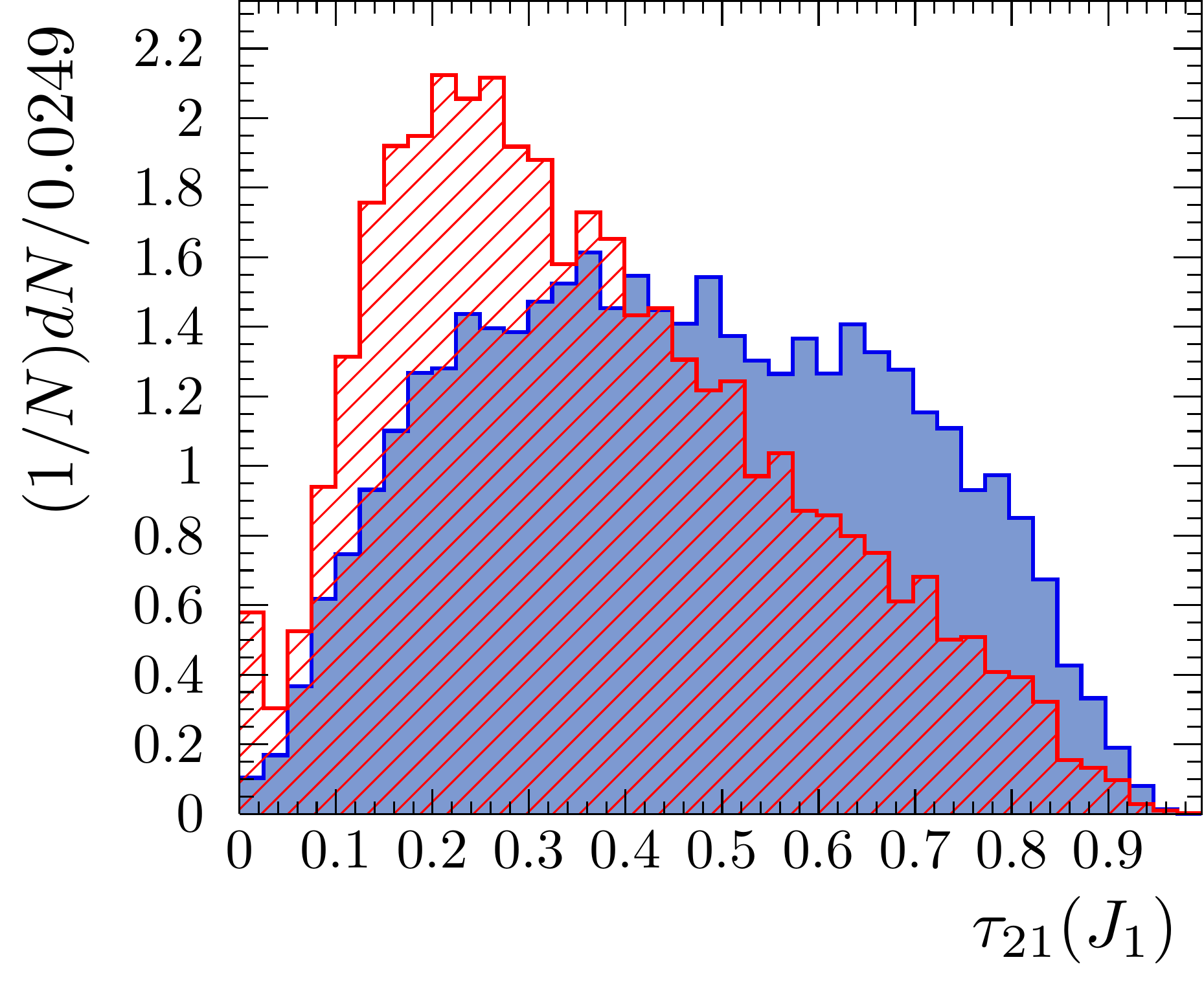}}\\
\caption{Normalized distributions of the additional input high level variables constructed for the fat-jets at the LHC ($\sqrt{s}=14$ TeV) used in the MVA for the signal (red) and the background (blue). Signal distributions are obtained for benchmark point {\bf BP1} and the background includes all the dominant backgrounds.}
 \label{fig:4_2}
\end{figure}

We use the following thirteen observables as input to BDT network. The normalized distributions of these input variables are shown in figure \ref{fig:4_1}, figure \ref{fig:4_2},  where the number on Y-axis represents the bin size.
\begin{itemize}
    \item Transverse momentum of leading fat-jet $P_T(J_0)$, figure \ref{fig:PT_J0}. 
    
    \item Transverse momentum of sub-leading fat-jet $P_T(J_1)$, similar figure not shown.
    
    \item The angular distance difference between two fat-jets $\Delta R(J_0,J_1)$, figure \ref{fig:Delta_R_J0_J1}
    
    \item The missing transverse energy $\slashed{E}_T$, figure \ref{fig:MET}
    
    \item The azimuthal angle difference between missing transverse energy and leading fat-jet $\Delta\phi (J_0,\slashed{E}_T) $, figure \ref{fig:Delta_Phi_J0_MET}
    
    \item The azimuthal angle difference between missing transverse energy and sub-leading fat-jet $\Delta\phi (J_1,\slashed{E}_T), $ figure \ref{fig:Delta_Phi_J1_MET}
    
    \item The effective mass of the process $M_{eff} = \sum_{vis} |P_T| + |\slashed{E}_T|$, shown in figure \ref{fig:Meff}

\item  The mass of leading fat-jet $M_{J_0}$ and sub-leading fat-jet $M_{J_1}$ are shown in figure \ref{fig:MJ0} and  figure \ref{fig:MJ1}, respectively. We used the pruned jet mass by applying the pruning method described in references \cite{Ellis:2009su,Ellis:2009me} to clean the softer and wide-angle emission. We first calculate $z = min(P_{Ti},P_{Tj})/P_{T_{i+j}}$ and the angular separation $\Delta R_{ij}$ between two proto-jets $i$ and $j$ at each step of recombination. Now, the softer proto-jet is discarded  if $z < z_{cut}$ and $\Delta R_{ij} > R_{fact}$ and $i$-th and $j$-th proto-jets are not recombined. Otherwise, $i$-th and $j$-th proto-jets are recombined,  and the procedure is repeated unless we remove all the softer and wide-angle proto-jet from the fat-jet. We have used a fixed $R_{fact}=0.5$ and $z_{cut}=0.1$ as suggested in reference \cite{Ellis:2009su}. 
    
    \item We use 2-prong discriminant energy correlation functions \cite{Larkoski:2014gra} 
    \begin{eqnarray}
    C_2^{(\beta)} = \frac{ e_3^{(\beta)}}{ (e_2^{(\beta)})^2}
    \end{eqnarray}
    where, $e_2^{(\beta)} = \sum_{1\leq i < j\leq n_J} z_i z_j \theta_{ij}^\beta$ and $e_3^{(\beta)} = \sum_{1\leq i < j < k\leq n_J} z_i z_jz_k \theta_{ij}^\beta \theta_{ik}^\beta \theta_{jk}^\beta$ are 2-point and 3-point energy correlation functions respectively. The $\beta$ represents the exponent.  Here $z$ is the energy fraction variable, and $\theta$ is angular variable. The distributions of $C_2$  for leading and sub-leading fat-jets are shown in figure \ref{fig:C22_J0} and \ref{fig:C22_J1} respectively. 
    
\item To reveal the two-prong nature of the fat-jet, we also use the N-subjettiness ratio \cite{Thaler:2010tr,Thaler:2011gf} 

\begin{eqnarray}
\tau_N^{(\beta)} = \frac{1}{\mathcal{N}_0} \sum\limits_i p_{i,T} \, \min \left\lbrace \Delta R _{i1}^\beta, \Delta R _{i2}^\beta, \cdots, \Delta R _{iN}^\beta \right\rbrace
\label{eq:nsub_N}
\end{eqnarray}
where, $\mathcal{N}_0=\sum\limits_i p_{i,T} R_0$ is the normalizing factor, $ R_0$ is the radius parameter of the fat-jet, N is the axis of the subjet assumed within the fat-jet and $i$ runs over the constituents of the fat-jet. We take the thrust parameter $\beta =$ 2 which gives more weightage to the angular separation of the constituents from the subjet axis. The distributions of N-subjettiness for leading and sub-leading fat-jets are shown in figure \ref{fig:Tau21_J0} and \ref{fig:Tau21_J1}. We choose {\tt One Pass $K_T$ Axes} for the minimization procedure. The N-subjettiness is slightly less powerful here because of high energetic fat-jets, wherein some events decay products become highly collimated and it is difficult to see a 2-prong structure.

\end{itemize}

\begin{figure}[t!]
	\centering
  \subfloat[]{\label{fig:Background_correlation}\includegraphics[scale=0.38]{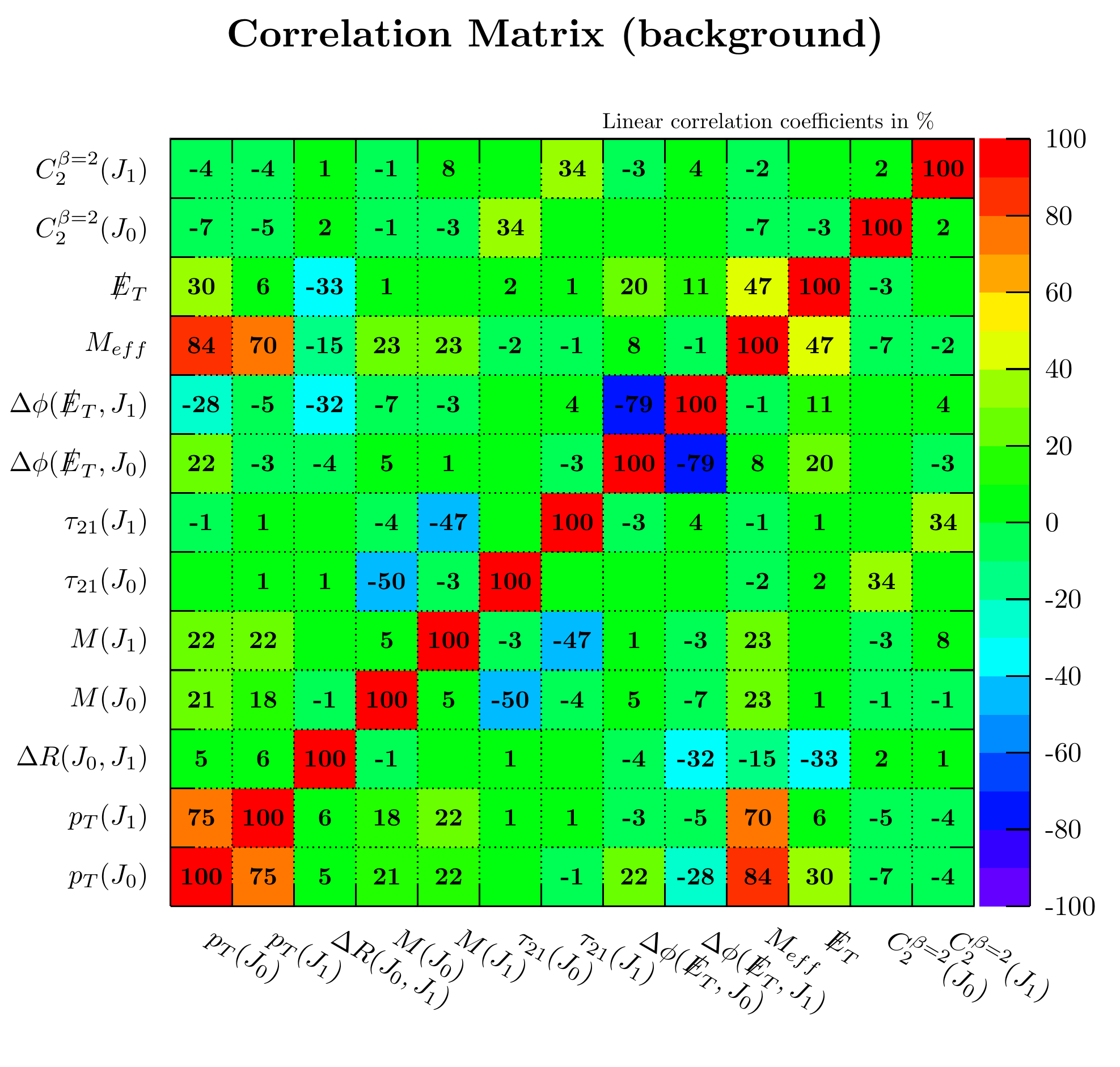}}~
  \subfloat[]{\label{fig:Signal_correlation}\includegraphics[scale=0.38]{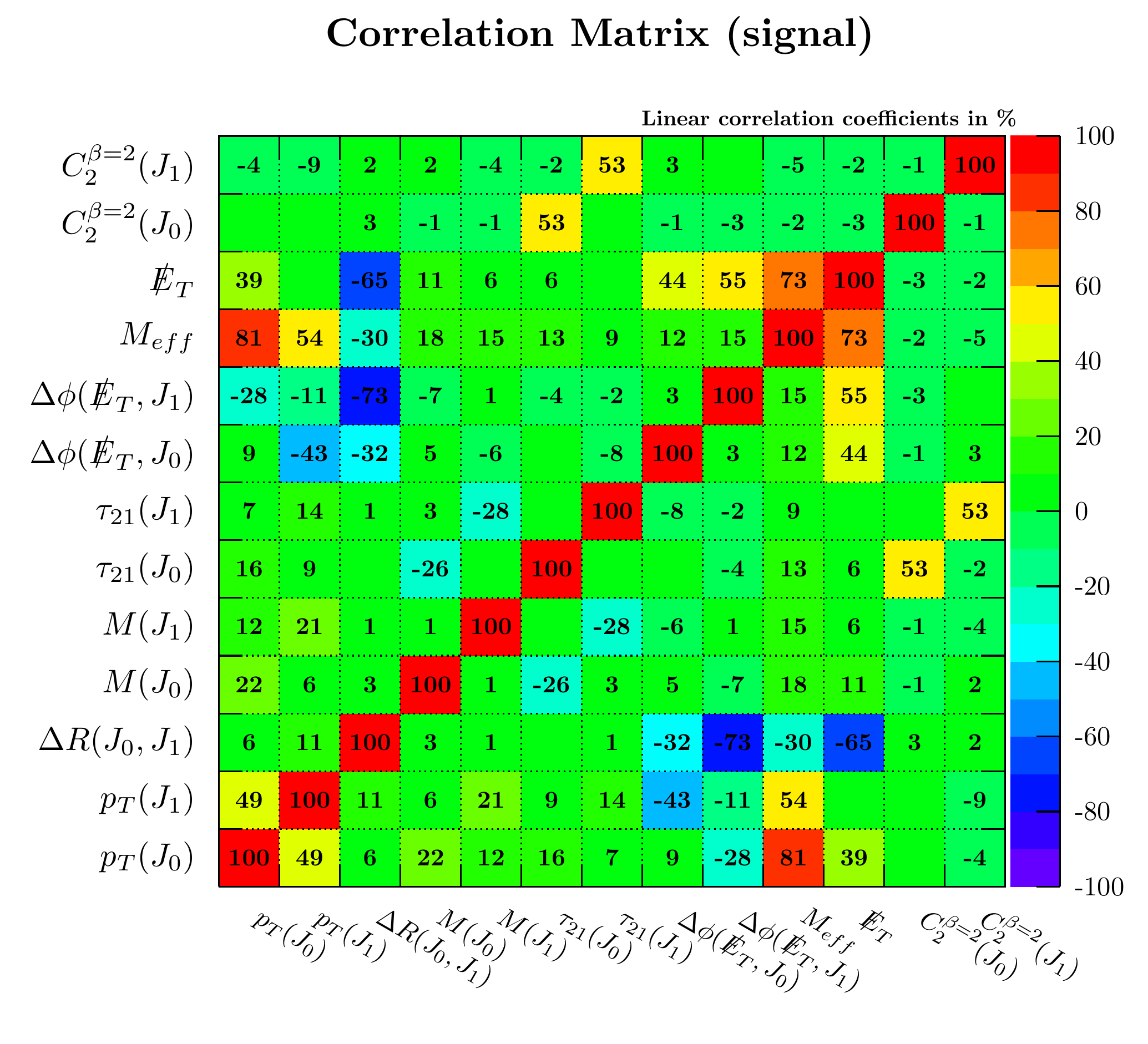}}
\caption{The linear correlations coefficients (in \%) for (a) signal and (b) background among different kinematical variables that are used for the MVA for benchmark point {\bf BP1}. Positive and negative signs of the coefficients signify that the two variables are positively correlated and negatively correlated (anti-correlated).}
	\label{fig:correlation}
\end{figure}

We calculate  the linear correlation $\rho$  between two  variable $X$ and $Y$ using the following equation

\begin{equation}
\rho(X,Y) = \frac{E(XY)-E(X)E(Y)}{\sigma(X)\sigma(Y)}
\end{equation} 
where $E(X), E(Y),$ and $E(XY)$  are the expectation value of the variable $X$, $Y$, and $XY$ respectively. Here, $\sigma(X)$ $\sigma(Y)$ represents the standard deviation of variable $X$ and $Y$ respectively.  Linear correlation among the variables plays a crucial rule to determine the information carried by the variable is unique or not. Most of the variables used in this study are highly uncorrelated with each other as shown in figure \ref{fig:correlation}. Here positive and negative sign of the coefficients signify correlation and anti-correlation with the other variable. Some set of variables like \{$P_T (J_0)$, $P_T (J_1)$, $M_{eff}$\} and \{$\Delta\Phi(\slashed{E_T}, J_0)$, $\Delta\Phi(\slashed{E_T},J_1)$\} show slightly high correlation for signal but have mild correlation in the background. This is mainly because of different kinematics of signal and background processes. Although one should use less correlated variables,  some variables with high importance are still used. This is mainly decided when the variable shows a different correlation for the signal and the background. These variables can have less correlation in different regions of phase space after the BDT applies the cuts. If the correlation is different for the signal and the background then a variable is selected/rejected depending on its importance.

\begin{figure}[thb]
\centering
\includegraphics[scale=0.50]{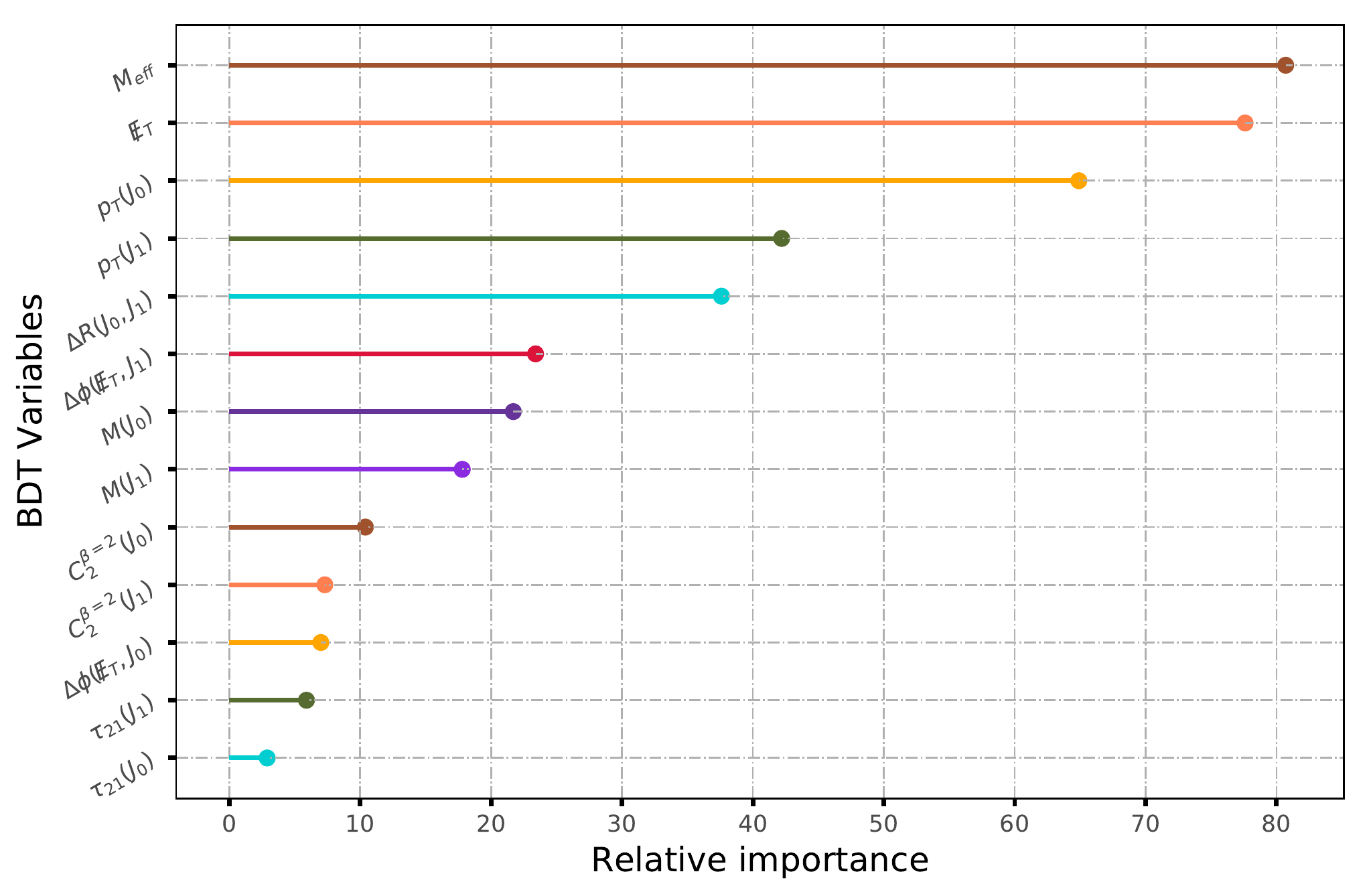}
\caption{ Kinematic variables used for our MVA and their relative importance. We obtain these using  numbers from the TMVA package for the benchmark point. Here, we show method unspecific relative importance.}
	\label{fig:Importance}
\end{figure}

We further show the method unspecific ranking (relative importance) for each
observable according to their separation in figure \ref{fig:Importance}. The separation in  terms of an observable $\lambda$ is defined as \cite{Hocker:2007ht}
\begin{equation}
\Delta_{(\lambda)} = \int \frac{(\hat{y}_s(\lambda)-\hat{y}_b(\lambda))^2}{\hat{y}_s(\lambda)+\hat{y}_b(\lambda)} d\lambda
\end{equation}
where $\hat{y}_s$ and $\hat{y}_b$  are the probability distribution functions for signal and background for a given observable $\lambda$ respectively. The limits of integration correspond to the allowed range of $\lambda$. 
Here $ \Delta_{(\lambda)}$ quantify discrimination performance of the observable $\lambda$. The separation $\Delta_{(\lambda)}$ ranges from 0 to 1. If $\Delta_{(\lambda)} = 0 (0\%) $ implies $\hat{y}_s(\lambda) = \hat{y}_s(\lambda)$, which means zero discrimination power of observable $\lambda$ and $\Delta_{(\lambda)} = 1 (100\%) $ corresponds to perfect discrimination power. 

After calculating the importance of variables, we divide the data set in two equal parts. One part of the data sample is used to train the BDT algorithm and the other part is used for the validation. The parameters used to train the BDT algorithm are shown in table \ref{tab:BDT_parameter}. 

\begin{figure}[tbh]
	\centering

   \subfloat[]{\label{fig:Train_test}\includegraphics[scale=0.38]{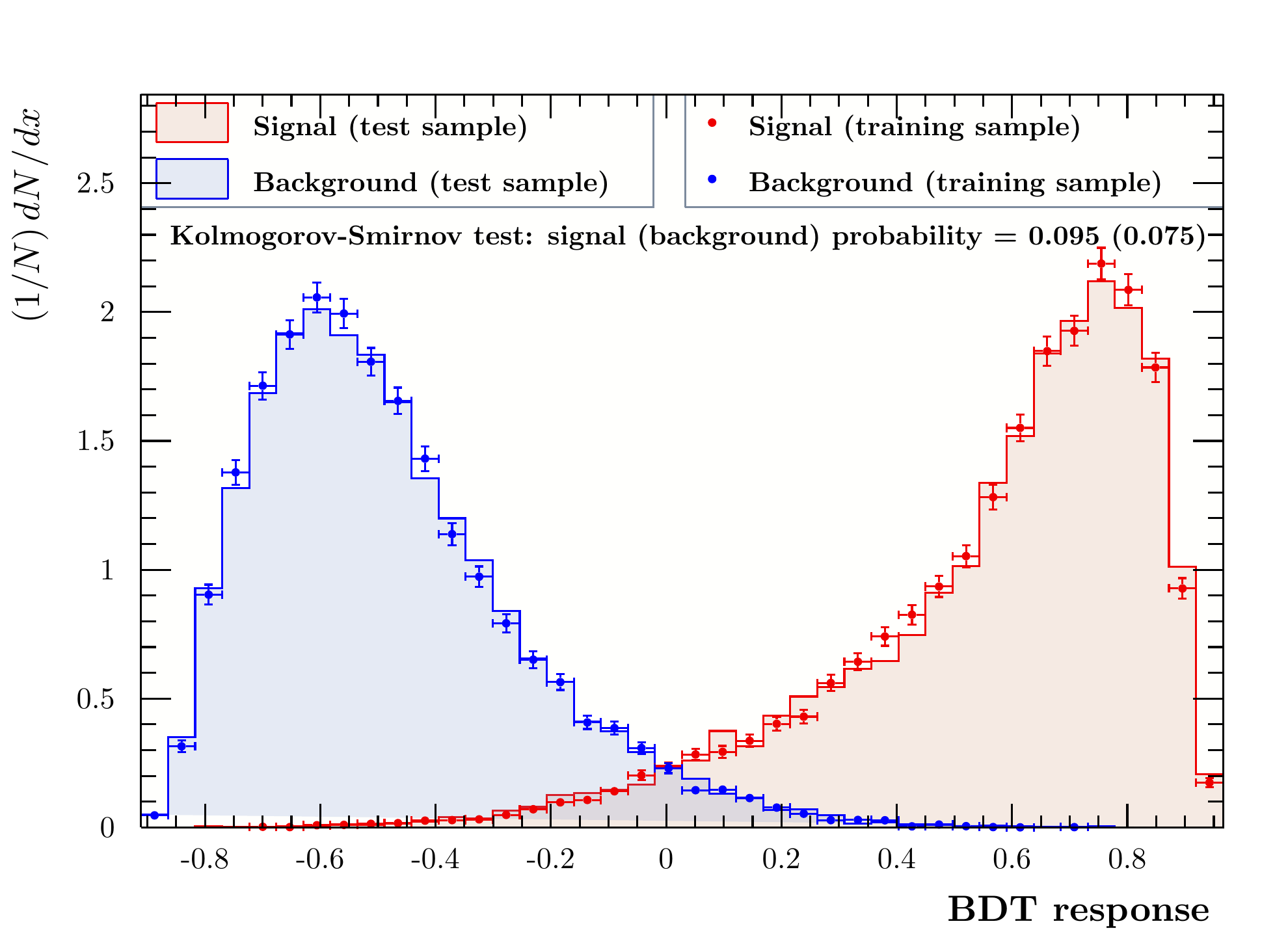}}~
      \subfloat[]{\label{fig:Significance}\includegraphics[scale=0.38]{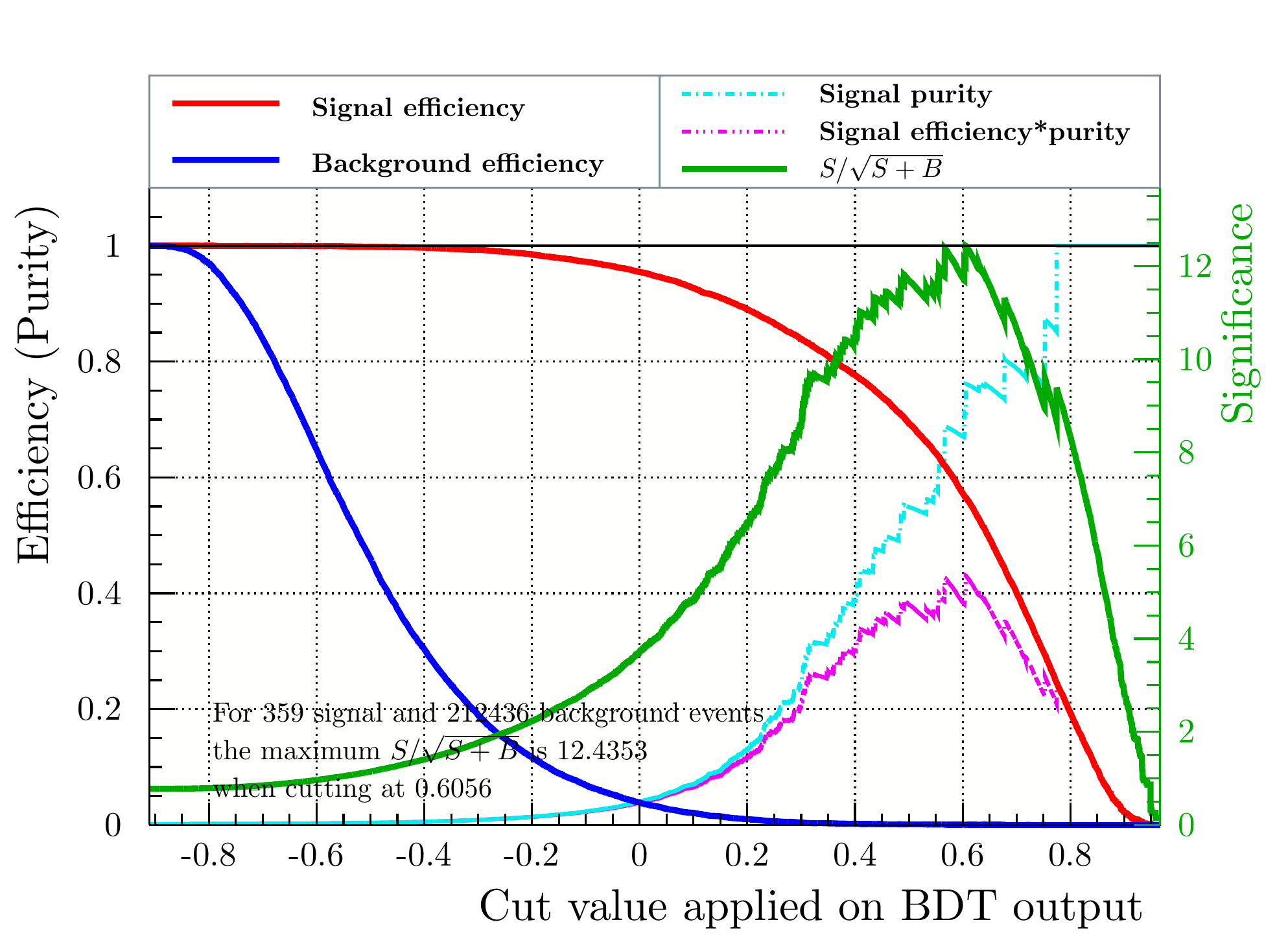}}
\caption{
(a) Normalized BDT response distributions for the signal and the background for the benchmark point {\bf BP1}.
(b) Cut efficiencies as functions of BDT cut values. All plots are evaluated for for benchmark point {\bf BP1} using integrated luminosity of $200$ fb$^{-1}$ at the 14 TeV LHC.}
	\label{fig:BDTresponse}
\end{figure}

\begin{table}[H]
	\centering
	\small
	\renewcommand{\arraystretch}{1.5}
	\begin{tabular}{|c|c|c|}
		\hline
		NTrees & 400 & Number of trees in the forest \\
		\hline
	 MaxDepth& 2 & Max depth of the decision tree allowed\\ 
		\hline
      MinNodeSize  & $5.6 \%$& Minimum \% 
            of training events required in a leaf node\\ 
		\hline
			BoostType & AdaBoost &  Boosting type for the trees in the forest\\
				\hline
AdaBoostBeta	& 0.5 & Learning rate for AdaBoost algorithm\\
		\hline
nCuts & 20& Number of grid points in variable \\

& & range used in finding optimal cut in
node splitting\\
\hline
\end{tabular}
\caption{Parameter used in BDT architecture}
\label{tab:BDT_parameter}
\end{table}
Results from BDT analysis considering one sample benchmark point (BP1) is demonstrated in figure \ref{fig:BDTresponse}. 
Kolmogorov-Smirnov probability for training and testing sample are shown to confirm that the network is not overtrained. The testing data fit well to the training data and the validation is shown in figure \ref{fig:Train_test}. The BDT is trained for each benchmark point separately. We apply the cut on BDT response and obtain the corresponding number of signal $\mathcal{N}_{S}$ and background $\mathcal{N}_{B}$.

\begin{table}[H]
	\centering
	\begin{tabular}{c|cccccc}
		\hline \hline
		BPs & $\mathcal{N}_S^{bc}$ & $\mathrm{BDT}_{opt}$ & $\mathcal{N}_{S}(\epsilon_S)$ & $\mathcal{N}_{B}(\epsilon_B \times 10^{4})$ & $\mathcal{N}_{S}/\sqrt{\mathcal{N}_{S}+\mathcal{N}_{B}}$ & $\mathcal{L}^{req}_{(5\sigma)} fb^{-1}$ \\ 
		\hline 
		BP1 & 359  & 0.60 & 202 (0.56) & 63 (2.9) &12.4 & 32.3 \\ 
		\hline
	    BP2 & 256  & 0.67 & 137 (0.56) & 50 (2.3)  &10.0 & 49.7\\ 
		
        \hline
        BP3 & 346  & 0.42 & 183 (0.52) & 49 (2.3)  & 12.0 & 34.5 \\ 
        \hline
        BP4 & 153  & 0.65 & 87 (0.56) & 15 (0.7) &  8.6 & 67.4\\ 
        \hline
        BP5 & 32   & 0.61 & 25 (0.78)&  51 (2.4) & 2.9 & 595.4\\ 
        \hline
        BP6 & 74  & 0.58 & 37 (0.50) & 42 (1.9)  & 4.2 & 283.2 \\ 
        \hline
        U1 &  266  & 0.57 &  149 (0.56) & 49 (2.3) & 10.6 & 44.4\\ 
        \hline
        U2 & 352  & 0.56 & 216 (0.61) & 41 (1.9) & 13.5 & 27.4\\ 
		\hline \hline 
		$\mathcal{N}_{\textrm{SM}}$ & 212436 & - & - & - & - \\ 
		\hline \hline
	\end{tabular} 
	\caption{Total number of signal events $\mathcal{N}_S^{bc}$ and background events $\mathcal{N}_{\textrm{SM}}$ before  utilising the optimum BDT criteria $\mathrm{BDT}_{opt}$ for an integrated luminosity of $200$ fb$^{-1}$ at the 14 TeV LHC. The number of signal and background events after the $\mathrm{BDT}_{opt}$ cut are
	denoted by $\mathcal{N}_S$ and $\mathcal{N}_{B}$ respectively. Here $\epsilon_S$ and $\epsilon_B$ represents the signal acceptance and background acceptance efficiency at the $\mathrm{BDT}_{opt}$ cut value. Finally, listed the statistical significance for an integrated luminosity of $200$ fb$^{-1}$ and also required luminosity for a five sigma discovery in case of each BP.}
	\label{tab:BDT}
\end{table}
 Finally we calculate the statistical significance using formula $\sigma=\mathcal{N}_{S}/\sqrt{\mathcal{N}_{S}+\mathcal{N}_{B}}$.
The cut value of BDT response is $\mathrm{BDT}_{opt}$, where the maximum significance is achieved. These steps were depicted in second plot for the sample benchmark point, as shown in figure \ref{fig:Significance}. Finally, the results for all benchmark points are displayed in table  \ref{tab:BDT}.

\subsection{Complementary signals at high energy and high luminosity upgrades of LHC at $\sqrt{s}$=27 TeV}

	\begin{table}[tbh]
		\begin{center}
			\begin{tabular}{|c|c|c|c|}
				\hline
				Channel & $\sqrt{s}=14$ ($\mathcal{L}=3 ab^{-1}$)  TeV & $\sqrt{s}=27$ TeV  ($\mathcal{L}=15 ab^{-1}$) \\
				\hline
				$(jj)(jj)$ &   4593  & 756177  \\
				$(jj)(ll)$ &    352  &  58011 \\
				$(ll)(ll)$ &    13 &    2126 \\
                $(jj)(jjl)$&     4  &  664   \\
                $(ll)(jjl)$ &    1 &  157 \\ 
				\hline
				
			\end{tabular}
			\caption{Number of events computed using $\sigma * BR$ for \textbf{BP1} at NLO for $\sqrt{s}=14$ ($\mathcal{L}=3$ab$^{-1}$) and 27 TeV ($\mathcal{L}=15$ab$^{-1}$) at LHC before analysis cuts are applied.}
	    	\label{tab:est}
			\end{center}
	\end{table} 
Semi-leptonic and leptonic channels with leptons inside the fat-jet, i.e, \textit{lepton-jets} are potential alternate channels to confirm the presence of the higgsino-like NLSP besides the hadronic channel. For example, the decay chain $\widetilde{\chi}_1^0 \rightarrow hh/ hZ , (h \rightarrow WW^*), (W \rightarrow j j, W \rightarrow l \nu) $ will give rise to an interesting signature of a lepton inside the fat-jet due to high boost of the Higgs. Note that a leptonic decay of the $Z$ boson would also lead to a pair of collimated leptons in the final state.  Therefore new signatures with lepton(s) inside jets such as $(jj) (jj),(jj) (ll)$, $(jj)(jjl)$ and $(ll)(ll)$ along with $\slashed{E}_T$ (where $l = e, \mu$) may serve as complementary signals to identify the current scenario. We estimate the number of events prior to signal analysis as summarised in  table \ref{tab:est} for $\sqrt{s}=14$ (27) TeV at 3 (15) ab$^{-1}$. We have used the \texttt{NNPDF} \cite{Ball:2012cx} parton distribution function to generate the signal events at $\sqrt{s}=27$ TeV and obtained the K-factors at NLO from \texttt{Prospino}\cite{Beenakker:1996ed,Plehn:2004rp,Spira:2002rd,Beenakker:1997ut,Beenakker:1996ch}.

From table \ref{tab:est} it is observed that the fully hadronic final state $(jj)(jj)$ is the best channel for discovery of the higgsino NLSP scenario over the other leptonic and semi-leptonic channels  due to  the dominant branching fraction into the hadronic channel. Although the number of events are expected to fall after all detector effects such as reconstruction efficiencies of the jets and leptons are taken into account. Further, signal selection criteria would also lead to reduction in the number of observed events. Therefore, at $\sqrt{s}=14$ TeV, only the fully hadronic channel is the best possible channel for discovery of the higgsino-NLSP scenario.  From section \ref{sec:analysis}, at $\sqrt{s}=14 $ TeV we see that the two fat-jet + $\slashed{E}_T$ final state can reach a mass range of $\simeq 2.4$ TeV- $3.0$TeV at an integrated luminosity, $\mathcal{L} = 200$ fb $^{-1}$.
Although the semi-leptonic channels $(jj)(ll)$ and $(jj)(jjl)$ can be interesting channels of discovery due to the presence of leptons in the final state, they have relatively fewer events at $\sqrt{s}=14$ TeV and are not expected to be significant after detector effects and signal selection efficiencies are taken into account. However such channels would possibly be discoverable at the high energy upgrade of the LHC at $\sqrt{s}=27$ TeV as shown in table \ref{tab:est}. The dilepton pair $(ll)$ arising from the decay of the $Z$ boson would also be an indicator of the composition of the NLSP since the $Z$ boson arising from the decay of the higgsino-like NLSP would be longitudinally polarised in the high energy limit where $\sqrt{s}>>m_Z$. On the contrary, a gaugino-like NLSP would give rise to a mostly transversely polarised $Z$ boson. Therefore, the presence of the longitudinal $Z$  boson  would be useful to ascertain the higgsino-like nature of the NLSP. Kinematic observables such as $\cos \theta^*$ and other variables derived therefrom are useful to explore the polarisation of the $Z$  boson as has been studied in \cite{Dutta:2019gox} for non-boosted topologies. We leave such studies using boosted techniques for a future work. In addition, channels including a lepton inside a jet, such as $(jjl)$ dominantly arise from the decay of the Higgs, $h \rightarrow WW^* \rightarrow jjl$ in the final state. It would be a useful indicator of the presence of a Higgs boson in the final state as opposed to a $Z$ boson and thereby affirming the higgsino-like composition of the NLSP.

\section{Distinction of Compressed and Uncompressed spectra}
\label{sec:distinct}
As the results suggest in table \ref{tab:BDT}, the signal yield for different compression is similar for a few benchmarks. It is important to compare the scenario of different compression scale. We define $\Delta M$ as compression scale, where  $\Delta M$ is the mass difference between the heaviest colour particle and the NLSP. $\Delta M$ varies from 56-190 GeV for the case of $\mathscr{C}$SUSY spectra while for uncompressed it is in between 500 - 2000 GeV. With $\widetilde{G}$ being almost massless and NLSP being in the range of (1-3 TeV) we expect that the decay product of NLSP will be sufficiently boosted in both the cases\footnote{Note that direct searches for the weakly interacting NLSP with a gravitino LSP already constrain the mass of such an NLSP to be heavier than 800 GeV \cite{Sirunyan:2018ubx,ATLAShiggsinoMSSM}.}. Hence both kinds of compression spectra satisfy the loose criteria of at least two fat-jet. 

\begin{figure}[!tb]
\centering
\includegraphics[width=9.6cm,height=6cm]{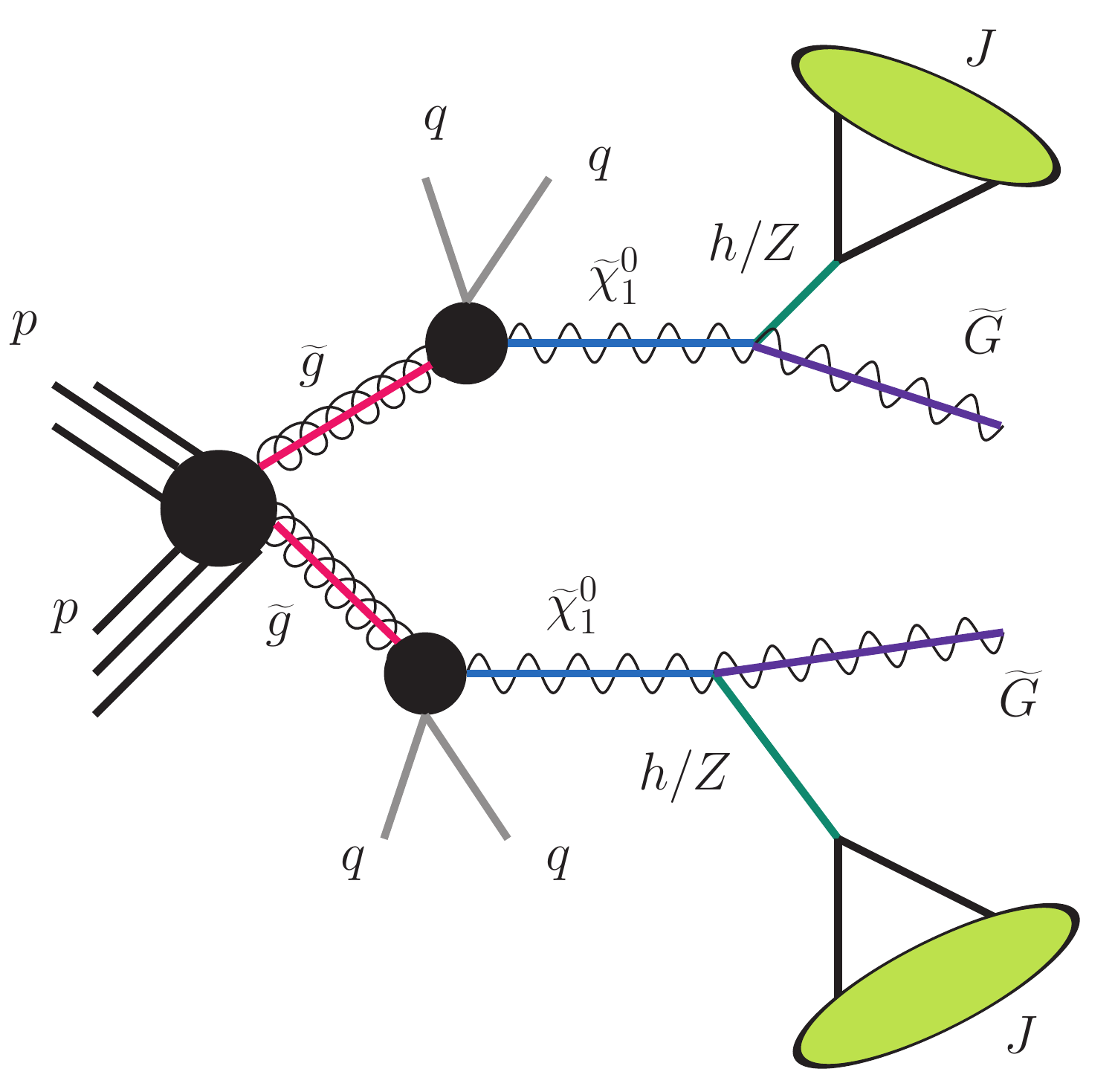}
\caption{ Representative diagram for the signal topology.}
	\label{fig:signaltopology}
\end{figure}
A large number of high $p_T$ jets are the result of the cascade decay in case of the uncompressed spectrum, whereas the compressed spectrum has very soft jet coming from the cascade decay. Using this information we design two new observables to distinguish these two spectra. To understand the construction of these observables the prototypical signal topology is shown in figure \ref{fig:signaltopology}.

We first define the anti-kT jet (\texttt{AK4}) of radius parameter R = 0.4 with $P_T$ = 20 GeV. Further, we identify these \texttt{AK4} jets ($j_k$) as ``\textit{unique jet}'' jets which are not the part of fat-jet ($J_i$) {\sl{i.e.}} $\Delta R_{J_i j_k}$ 
between the reconstructed fat-jet and a \texttt{AK4} jet is greater than 0.8 hence \textit{unique jets} are well separated from the fat-jets\footnote{In this analysis, we consider two different classes of jets. \texttt{AK4} has characteristics and properties such as cone-like regular jet shape which makes it preferable for experimental use, both for jet energy calibration and subtraction of underlying events and pileup. Hence for all small-radius jets, we decided to consider the same. While the CA8 fat-jets are constructed to be used to study the sub-jet structure and variables in the pruning of such jets.}.  The origin of unique jets is primarily from cascade decay hence they can be identified in a small radius jet.

\begin{figure}[!tb]
	\centering
	\subfloat[]{\label{fig:Z1C1_U1}\includegraphics[scale=0.5]{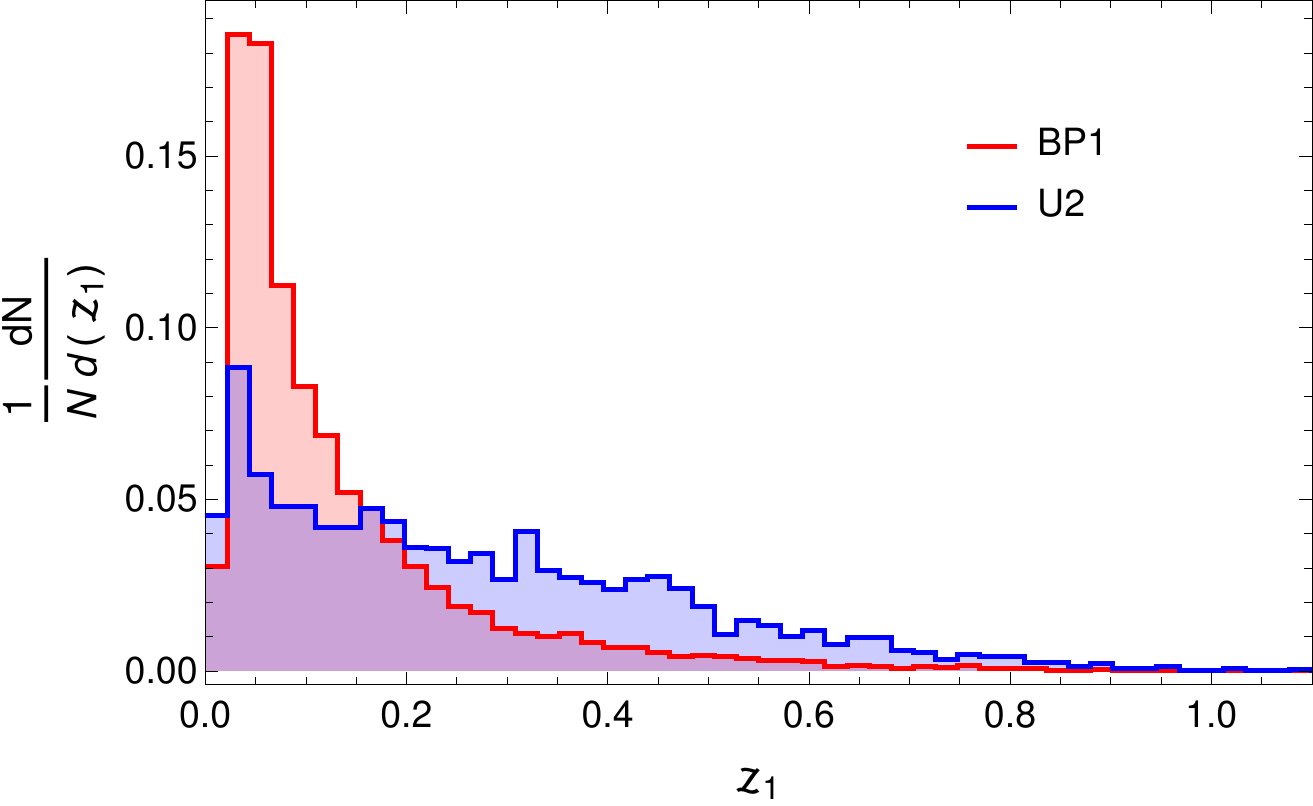}}~
	\subfloat[]{\label{fig:Z2C2_U2}\includegraphics[scale=0.5]{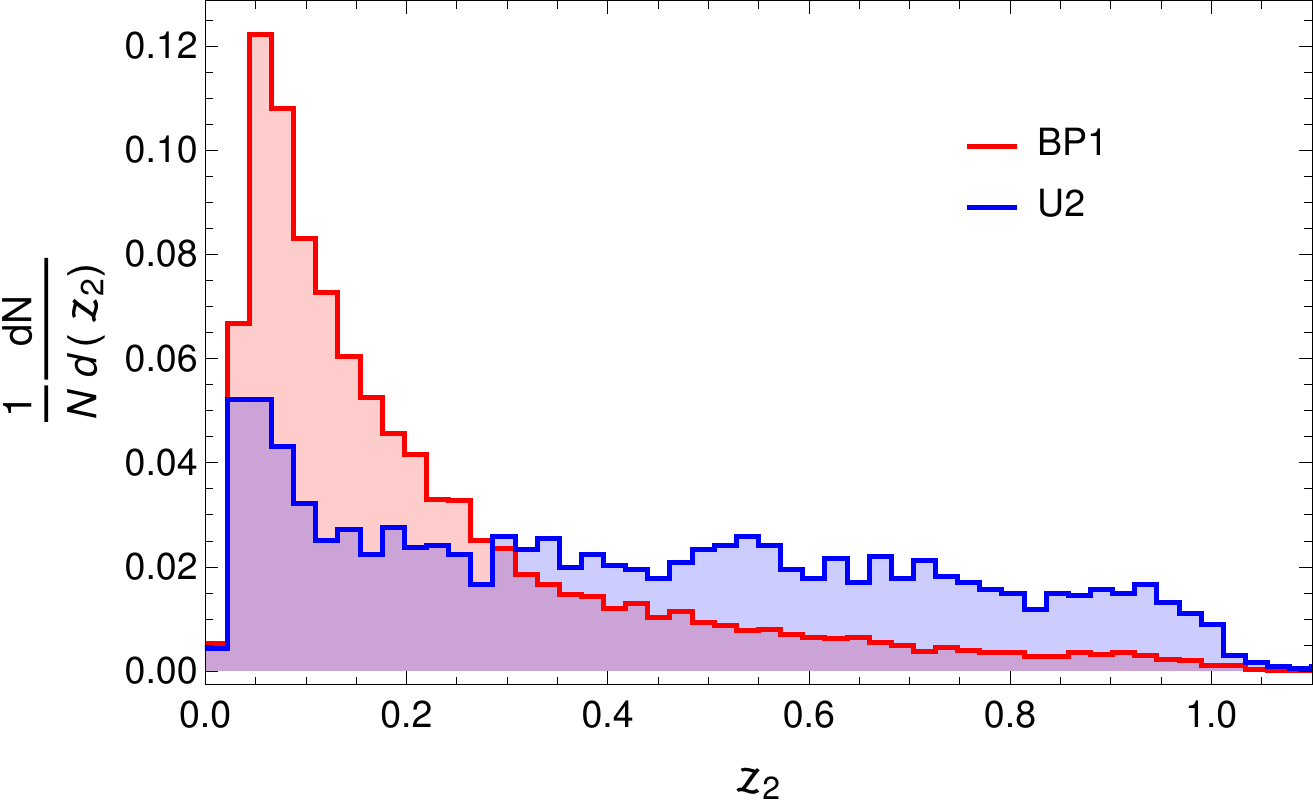}}\\
	\caption{Normalized distributions of new kinematic variable $ {\cal Z}_1$ and $ {\cal Z}_2$  for the discriminant of compressed and uncompressed spectra.}
	\label{fig:C1_U1}
\end{figure}

\begin{itemize}
 \item The first observable is defined as the ratio of $P_T$ of leading unique \texttt{AK4} jet by the $P_T$ of leading fat-jet, written as 
 \begin{equation}
 {\cal Z}_1=\frac{P_T{(j_0)}_{unique}}{P_T(J_0)}
 \end{equation}
 \item Similarly, we define another variable as the ratio of $P_T$ of leading unique jet by the $P_T$ of sub-leading fat-jet, written as
 \begin{equation}
 {\cal Z}_2=\frac{P_T{(j_0)}_{unique}}{P_T(J_1)}
 \end{equation}
 \end{itemize}
 The distribution for these variables are shown in figure \ref{fig:Z1C1_U1} and \ref{fig:Z2C2_U2}  respectively. These distributions are plotted with the selected events after the BDT analysis. Evidently, both variables can capture significant information about the compression of the spectrum. The $ {\cal Z}_1$ and $ {\cal Z}_2$ both have significant contribution at smaller value for  {\bf BP1} (compressed case) compared to a relatively flat distribution in {\bf U2} (uncompressed case).  As expected,  $p_T$ of the leading unique jet is less in case of compressed than in the case of uncompressed spectra and these variables can be used as powerful discriminators in hadronic final state studies of $\mathscr{C}$SUSY.

\section{Summary and Conclusions}
\label{sec:sumcon}
With no clear indication of new physics yet at the LHC, compressed mass spectrum gained significant limelight as a possible explanation for the elusive nature in the realisation of new physics.  In this work, we consider a compressed SUSY scenario, where both coloured and electro-weak new physics sectors are sitting at multi-TeV scale in the presence of a light gravitino as dark matter candidate. The lightest neutralino, which is also the natural NLSP candidate in phenomenological MSSM, decays into the gravitino together with Higgs or $Z$-boson. 
A large mass gap between them invariably produces a significantly boosted boson. 
Recognising the fact that its hadronic decay can form boosted fat-jet objects opens up an intriguing new possibility. This new channel can be beneficial contrary to looking through the typical leptonic search which is in any case expected to be suppressed by small branching ratio, or reconstruction efficiency at a high $p_{T}$. Moreover, reconstructed fat-jets can still carry the characteristics of the parent particle in their masses and substructures. The present analysis exploits such properties to counter the extensive background coming from  QCD jets.  With multiple observables, including pruned fat-jet masses, energy correlation functions as well as N-subjettiness, we demonstrate the full potential of jet substructure by using a dedicated multivariate analysis. The LHC sensitivity can be improved substantially that most of the constructed benchmark points can be explored with an integrated luminosity of 200 $fb^{-1}$ at the 14 TeV LHC. One can exclude masses up to 3.2 TeV at $\mathcal{L}=3000$ fb$^{-1}$, with a $3.2\sigma$ signal significance achievable for a compressed spectrum similar to \textbf{BP6} ($\Delta M \simeq 60$ GeV).

At this point, it is worth mentioning that an uncompressed scenario can produce characteristically different signature. We constructed new observables in our present framework sensitive to the compression of our model.
New possible leptonic and semi-leptonic signatures are also proposed which would be observable at a high energy and high luminosity upgrade of the LHC at $\sqrt{s}=27$ TeV.

\section{Acknowledgement}
\label{sec:ack}
 The work of AB and PK is supported by Physical Research Laboratory (PRL), Department of Space, Government of India and the computations were performed using the HPC resources (Vikram-100 HPC) at PRL. JD, BM and SKR acknowledges support from the Department of Atomic Energy, Government of India, for the Regional Centre for Accelerator-based Particle Physics (RECAPP), Harish-Chandra Research Institute. JD acknowledges support by the Deutsche Forschungsgemeinschaft (DFG, German Research Foundation)under Germany's Excellence Strategy EXC 2121 "Quantum Universe"- 390833306.
\providecommand{\href}[2]{#2}
\addcontentsline{toc}{section}{References}
\bibliographystyle{JHEP}
\bibliography{ref}

\end{document}